\documentclass[preprint,floatfix,aps,pre,amsfonts,amsmath,amssymb,superscriptaddress,showpacs]{revtex4-1}

\usepackage{geometry}
\usepackage[english]{babel}
\usepackage{lmodern}

\usepackage{epsfig}
\usepackage{graphicx}
\usepackage{caption}
\usepackage{subcaption}
\usepackage{color}
\usepackage{xcolor}

\usepackage{amsmath,amssymb}
\usepackage{array}
\usepackage{units}
 \usepackage{relsize}

\usepackage{natbib}

\begin{document}
\pdfoptionpdfminorversion=7

\title{Small- and Large-scale Characterization and Mixing Properties \\ in  a Thermally Driven Thin Liquid Film }

\author{Michael Winkler }
\affiliation{ University of Potsdam, Institute for Physics and Astrophysics, Potsdam, Germany }

\author{Markus Abel}
\affiliation{ University of Potsdam, Institute for Physics and Astrophysics, Potsdam, Germany }
\affiliation{ Ambrosys GmbH, Potsdam, Germany}

\begin{abstract}
Thin liquid films are nanoscopic elements of foams, emulsions and suspensions, and form a paradigm for nanochannel transport that eventually test the limits of hydrodynamic descriptions. Here we use classical dynamical systems characteristics to study the complex interplay of
thermal convection, interface and gravitational forces which yields
turbulent mixing and transport: Lyapunov exponents and entropies.
We induce a stable two eddy convection in an extremely thin liquid film by applying a
temperature gradient. Experimentally, we determine the small-scale dynamics using
the motion and deformation of spots of equal size/equal color, we dubbed that technique
``color imaging velocimetry''. The large-scale dynamics is captured by encoding
the left/right motion of the liquid directed to the left or right of the separatrix
between the two rolls. This way, we characterize chaos of course mixing in
this peculiar fluid geometry of a thin, free-standing liquid film.
\end{abstract}

\pacs{47.52.+j,47.55.P-,68.15.+e,47.27.wj,47.55.pb,47.27.-i,82.70.Rr,47.51.+a}

\maketitle

\section{Introduction}
\label{sec:intro}

Chaotic or turbulent mixing is essential for many industrial processes, so a profound understanding is good for applications.
However, despite the basic mechanisms for mixing in dynamical systems \cite{ottino1989kinematics,doering1995applied} are well understood,
mixing characterization in experiments may be hard due to the real-world applications. Here, the restrictions are finite time,
finite length and complex geometries.
The system under consideration is an aqueous thin film, driven thermally. The film is almost two-dimensional and
can show a thick phase (several $\mu m$) and a thin one ($\sim 10-50\,nm$), both immiscible due to the forces separating  them, similar to
bubbles (the thin phase) in a 3D liquid phase. The thinning speed of the film, i.e. the transition of the whole film
to the thin phase depends primarily on the mixing of thin and thick phase. Without mixing, such a film typically
undergoes thinning within several hours, with thermal driving, a flow is established that mixes the phases and
leads to a thinning in a time of the order of seconds. Consequently, it is of high interest to understand this
mixing and eventually quantify it. In the above mentioned article \cite{winkler2013exponentially} the basic physics
are explained. Here, we give a quantitative analysis of the mixing properties of the experiment in terms of
Lyapunov exponents and  entropies. We first discuss briefly the properties of thin liquid films, followed by an explanation of
our approach to the characterization of mixing for this highly sophisticated system.

Thin film dynamics is governed by \textit{gravitational}, \textit{capillary} and
\textit{interfacial} forces, where the latter are  specified in the \textit{disjoining
pressure}. Combining long- and short-range molecular forces:
electrostatic, van der Waals (vdW), and steric forces
\cite{derjaguin1989theory,oron1997long}, the disjoining pressure depends
strongly on the distance between the interacting surfaces. Whereas films on
substrates are established in industry and research \cite{reiter1998artistic},
freestanding thin liquid films still provide a challenge in experiments and theory alike.
Consequently, the study of foam films is central to current scientific
activities, e.g. \cite{kellay2011turbulence,yunker2011suppression,davey2010enantiomer,
Prudhomme-Khan-96,exerowa1998foam,vermant2011fluid}.
We contribute by presenting a way to quantify mixing of
\textit{vertically oriented, freestanding, thermally forced, nonequilibrium}
foam films.

As a result of the aforementioned force balance two stable equilibria may occur, depending on
the chemical composition of bulk solution and chosen surface active agents
(surfactants): Common Black Films (CBF) with a thickness of more than
\unit[10]{nm} are formed when electrostatic interactions balance the dominant
van der Waals force, and, of course, gravity and capillarity
\cite{derjaguin1989theory,Verwey-Overbeek-48}; Newton Black Films (NBF) are
stable  with  a thickness of less than \unit[10]{nm}, due to repulsive short
range steric forces \cite{jones1966stability,israelachvili1991intermolecular}.
In this study we will focus on films in their transient phase before reaching
equilibrium with a typical thickness in the range of $0.1 - 1 \,\mu m $. The
effect of additional forces has been studied by a several authors, mostly for
micrometer-thick systems \cite{zhang2005velocity,seychelles2008thermal}.

Quantifying the degree of complexity of an evolving system is an ubiquitous
problem in natural science \cite{badiicomplexity}. We will focus on two ways of
measuring dynamical complexity: the metric or Kolmogorov-Sinai entropy $h_{KS}$,
which measures the rate of information production as a fluid particle evolves
along a pathline, and the Lyapunov exponents (LEs), which give the rates at which
nearby fluid paths diverge.
The principle concept of the KS-entropy is very natural, as the information
contained in the time evolution is a characteristics of the underlying dynamics,
cf. Brudno's theorem \cite{batterman1996chaos}.

From data, one can obtain an estimate by studying the symbolic dynamics acquired by assigning
different symbols to different ``cells'' of a finite partition of the phase space.
The probability distribution of realized sequences (words) is a signature
dynamical evolution. The average information gain is obtained by comparing
sequences of length m and m+1, in the limit of large m: letting the length of
the words, m, to infinity and the partition diameter to zero, one obtains
the KS-entropy, which is often used as a ``measure of complexity'' of a system.

Lyapunov exponents $\lambda$ characterize the exponential divergence of nearby trajectories, typical for
chaotic systems \cite{ott2002chaos,kantz2004nonlinear}. One can alternatively state that LEs characterize the
sensitivity of a system to initial conditions $X_i$.
They are related to the KS-entropy $h_{KS}$ by the Pesin formula:
\begin{equation}\label{Pesin formula}
  h = \sum_{i=1}^{n}\lambda_i > h_{KS}\;.
\end{equation}
Positive Lyapunov exponents indicate that solutions diverge exponentially on
average, negative ones indicate convergence. They are computed from time series
using embedding techniques \cite{kantz2004nonlinear}, in our case, we can use the
spatial information directly and use  instead stretching and folding of a fluid
area to estimate  the  local dynamical characteristics directly from the experiment.

\section{The Experiment}
\label{sec:experiment}

In this section, we describe in detail the experiment, the flow structure which is crucial for
the analysis. Then, we explain the data analysis we applied, first for the LEs, then for the entropies.
In principle, we follow complementary approaches: the LEs are computed from microscopic information,
whereas the entropies are determined from the large-scale circulation.
Ideally, both quantities should coincide, to be tested by Pesin's formula. We will show the results
below.

\subsection{Setup}
\label{sec:setup}

\begin{figure}[ht]
\begin{center}
 \includegraphics[width=0.5\textwidth]{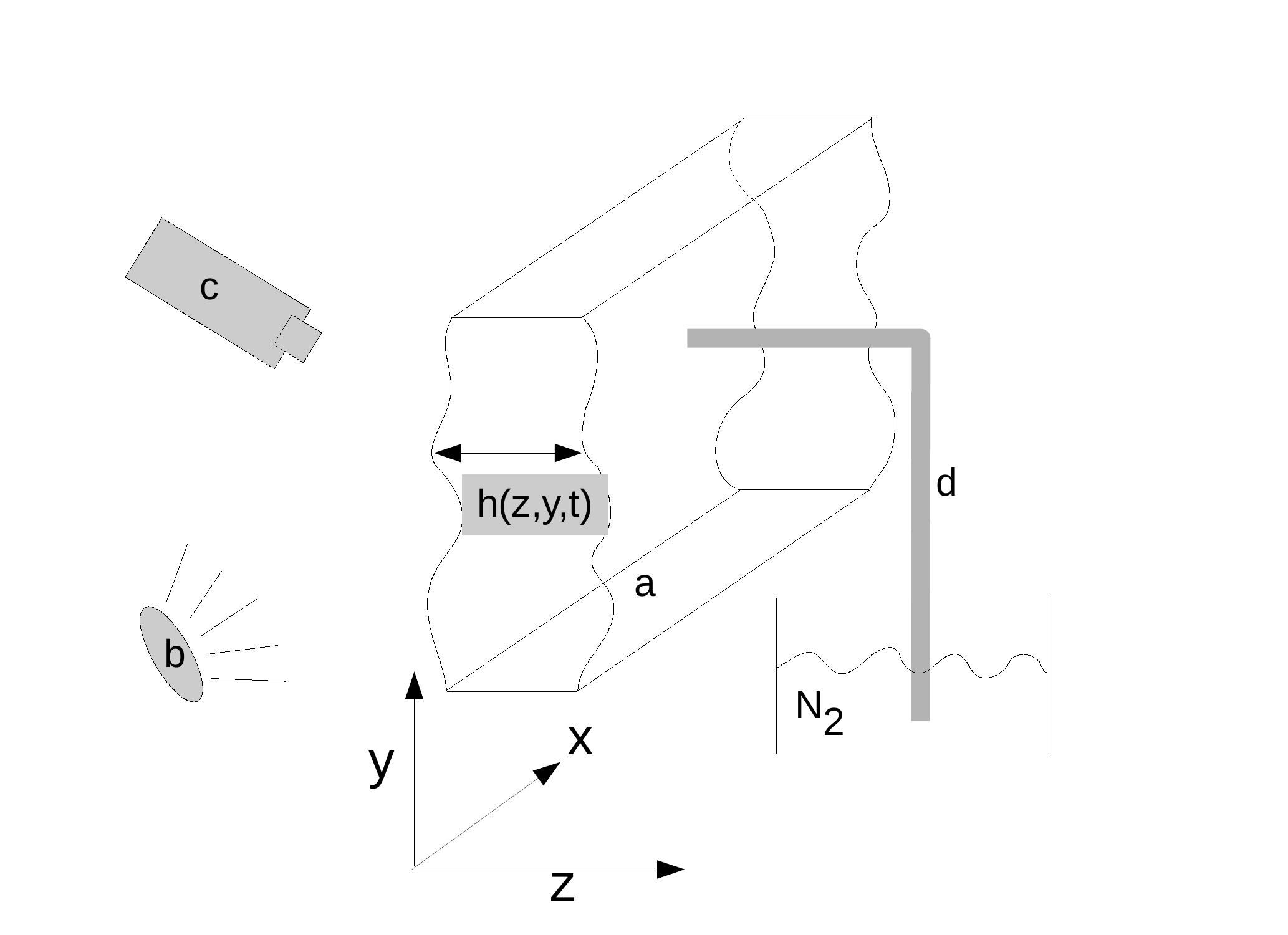}
\end{center}
 \caption{Setup of the experiment. (a) schematic sketch of the free-standing
thin film, (b) light source, (c) camera, (d) cooling rod with liquid nitrogen
reservoir.}
\label{fig:setup}
\end{figure}

The experimental setup consists of a vertical rectangular aluminium frame
with rounded corners, \unit $45 \times 20${ mm}, enclosed by an
atmosphere-preserving cell with a glass window for video recording,
cf.~Fig.~\ref{fig:setup}. Thermal forcing is effected by inserting a cooled
copper needle (radius \unit[1]{mm}) at the film center
(\unit[$T=-169$]{$^\circ$C}), the needle enters the cell through a fitting hole.
Ambient temperature was constant at \unit[20]{$^\circ$C}, the corresponding Rayleigh
number $Ra\sim 10^6$, such that the flow is clearly turbulent. Please note
that for very thin films it is not clear how viscosity changes with
the film thickness, since surface forces may matter. We are not aware of any investigations
in that direction, and will not discuss this further. The given number denotes an order of magnitude, such that only
dramatic changes in viscosity are relevant, e.g. if the film has a transition to thin black film
and changes structure, too. The temperature across the film
(z-direction, cf. Fig.~\ref{fig:setup}) is approximately constant, and the
Marangoni number $Ma\simeq 0$ .
The solution from which the liquid film was drawn consists of
the surfactant n-dodecyl-$\beta$-maltoside ($\beta\mathrm{-}C_{12}G_2$),
prepared with filtered deionized water and stabilized with 25~$\%_{\rm vol}$
glycerin \cite{stoeckle2010dynamics}.
The thin liquid film is illuminated with  a diffuse broad spectrum light source
and its reflection is captured by a high speed camera.
Our vertically oriented foam film is produced initially thick (500 -- 5000 nm)
by pulling it with a glass rod from the reservoir. Quickly, a wedge-like
profile develops with BF in a small region, with a sharp horizontal boundary
towards the thick film below, cf. Fig.~\ref{fig:snapshot}.

The interference of incident and reflected light yields a striped pattern,
which can be used to infer the film thickness. Each color cycle (red $\rightarrow$ blue) corresponds to
multiples $n$ of the smallest negative interference condition
$(2\cdot n+1) \lambda \eta = 4 h \cos \Theta$, where the refraction index,
$\eta$, is
assumed to be temperature-independent; $\Theta$ is the angle of incidence
\cite{atkins2010investigating}.
The velocity is measured by  color imaging velocimetry (CIV)
\cite{winkler2013mixing} with $u \sim  0.02\,\unitfrac{m}{s}$.

For ideal 2D flows -turbulent or not- surface forces are neglected. This is
reasonable for thick films. In our case, a full description involves all
occurring forces \cite{winkler2013exponentially}. For the Rayleigh number we are working with, $Ra \sim 10^6$, other forces
can be neglected to first order. Surfactants
stabilize the film, which is confirmed by our experiment, even if the film is heated
(giving rise to thermal fluctuations) and if the entire film is black, and of
the thickness of a few nanometers (giving rise to sensitivity to other molecular forces).
The turbulence generated by the cold needle is observable with the
naked eye, a snapshot  shown in Fig.~\ref{fig:snapshot}.

\subsection{Flow Description}

This publication presents experimental work and subsequent data analysis,
thus we do not discuss the equations of motion of the film,
but refer to recent work and reviews for the suiting equations in our situation
\cite{Kruesemann-2012,oron1997long,erneux1993nonlinear}.
However, we now discuss the observations in order to give the reader an impression
about the flow we  consider in terms of dynamical systems properties.

\begin{figure}
  \begin{minipage}[c]{0.5\textwidth}
   \includegraphics[width=0.47\textwidth]{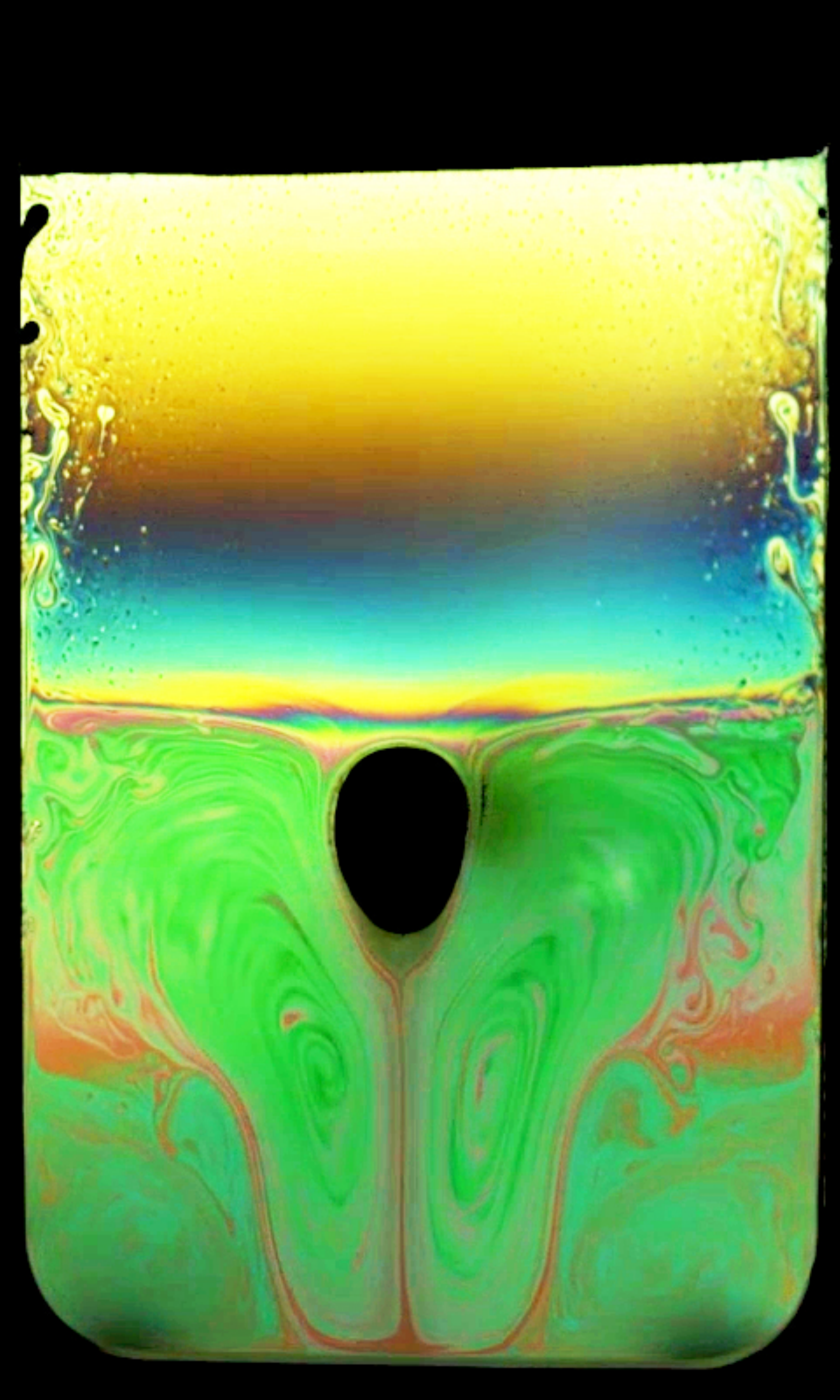}
    \includegraphics[width=0.47\textwidth]{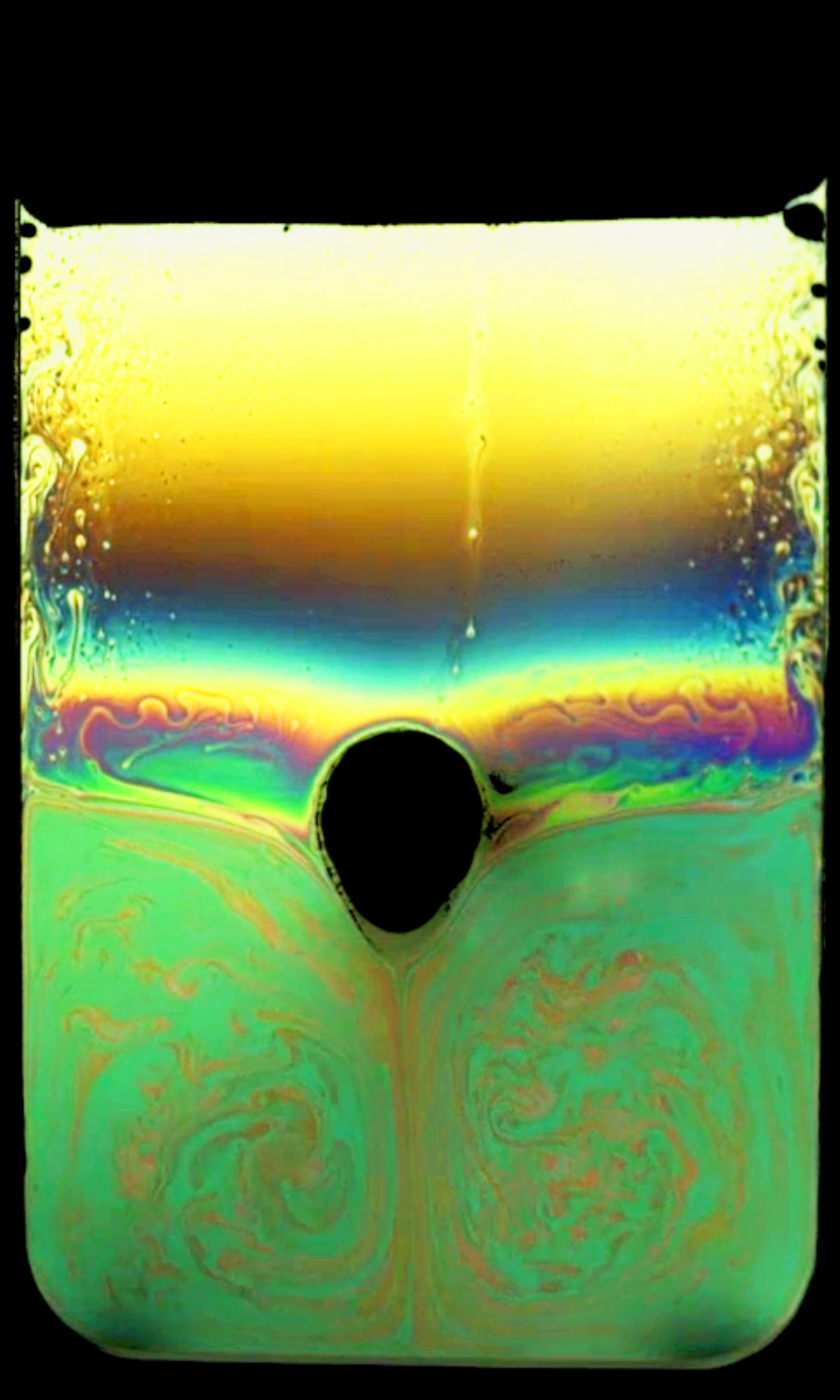}
  \end{minipage}\hfill
  \begin{minipage}[c]{0.5\textwidth}
    \caption{Snapshot of a turbulent convecting thin film. An image of the
    non-moving film has been subtracted, the frame and frozen center region
    including the cooling tip are not shown.
    On the sides, the typical meandering of rising thin/sinking thick film is
    observed. On the top, a layer of black film has formed which is transported
    downwards.
    Thickness can be read from the colors. Left: The convection is not yet fully
    developed, such that ``corner'' vortices are observed; they disappear after
    transients. Right: Turbulent convection has fully developed and thin film is
    transported and mixed quickly by the convection.}
  \label{fig:snapshot}
  \end{minipage}
\end{figure}

We use a point thermal force to drive our system, such that we have two
rolls, one on each side of the cooling rod, cf. Fig.~\ref{fig:snapshot}.
The cold rod drives a stable two-eddy convection with a jet of fluid in the
center as the driving mechanism and separation at the same time.
The deflection of the jet at the lower frame
border is sequential and non-deterministic. For the analysis of the large-scale flow,
the alternation between the left and the right eddy/vortex was visually tracked, recorded during a long time interval
and converted into a binary time series as detailed in section \ref{sec:dataanalysis}. Further information on this technique is available in our previous works \cite{winkler2013exponentially,winkler2011droplet,winkler2013mixing}.

\begin{figure}%
\centering
\includegraphics[draft=flase,width=0.35\columnwidth]{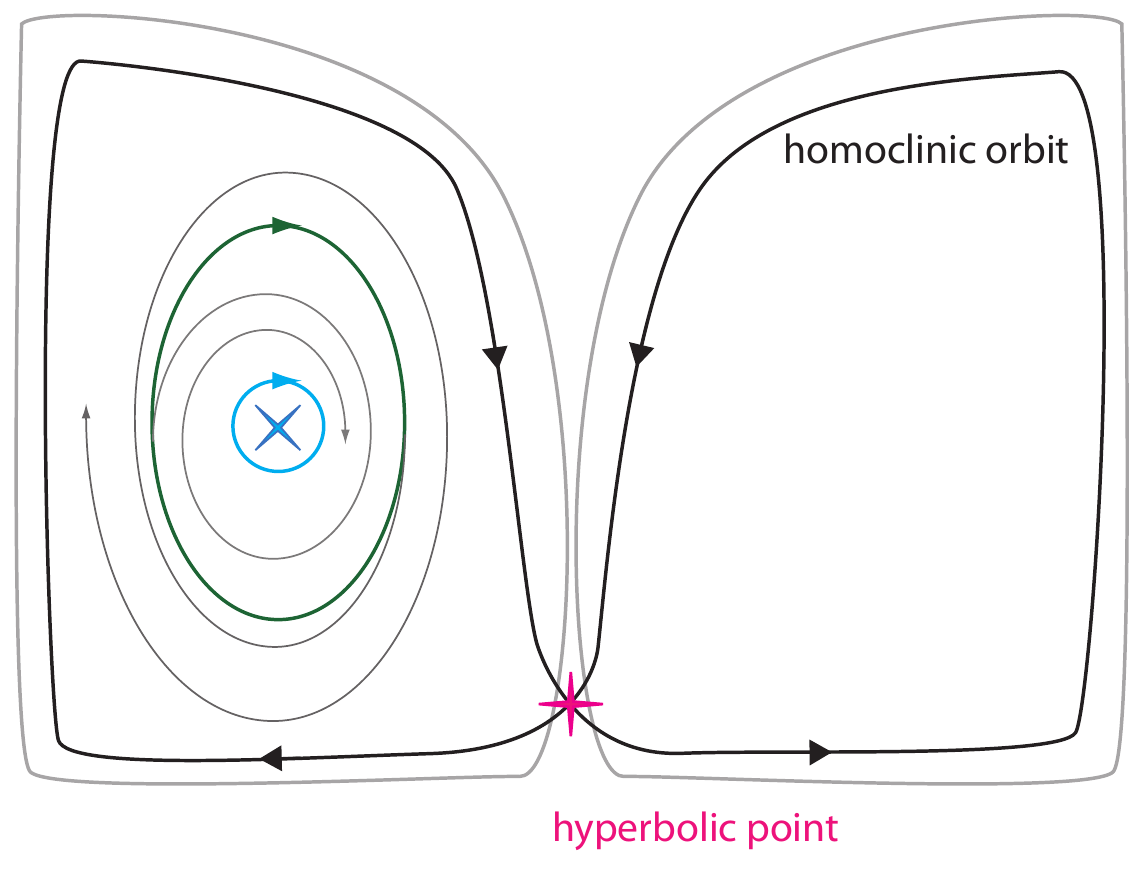}%
\caption[Orbit schematic for the presented experiment]{Orbit schematic for the
presented experiment (only the right half of the
convection zone is shown in detail). The light gray outer line represents the
shape of the two convection zones in the experiment. The gray traces represent
orbits moving away or towards the center of convection. The elliptic orbits are
marked with a light blue circle around the stationary elliptic point (blue
cross) in the center of the convection zones. Green marks the limit between
converging and diverging orbits.}
\label{fig:orbits_real}%
\end{figure}

In an abstract way, our flow can be described by geometry (boundary conditions) and four fixed points:
the flow is bounded by the enclosing rectangle, which in the upper region is given
by the black film region. Boundary conditions are complicated, in a first step they are assumed no-slip.
Then, we have the centers of the convection rolls, which are elliptic fixed points
There are two hyperbolic points at
the bottom and directly below the lowest part of the ice surrounding the
cooling needle and a separatrix running top-down (cf.
Fig.~\ref{fig:orbits_real}), whose connection form a separatrix.
This description holds on short time scales, for long times, we must consider that the whole
system is in a transient state which, however evolves on much larger time scales, such that
our considerations are reasonable. A look at corresponding video material will
clarify this characterization.

We want to characterize the flow, based on measurements. Key characteristics for mixing
and flow in general are stretching or folding of the liquid filaments.
We consider respective orbits of fluid elements advected by the two convection rolls, which
are stably positioned below the cooling rod. One key question regards the mixing  inside the rolls,
the other one mixing between the rolls. In contrast to other systems, as e.g. the blinking
vortex \cite{ottino1989kinematics}, the vortices are maintained and mixing happens across the
separatrix due to small differences in the middle downward flow.
Conceptually, we can use the basic ideas of twist maps, as one common example for
reduce dynamical systems showing mixing. I.e. we study the left-right transport, i.e.
the mixing between the two rolls.

Below, in Sec.\ref{sec:dataanalysis}, we explain how typical stretching rates
are extracted from the experiment, as described in more detail in
\cite{winkler2013mixing}. We recall the procedure briefly here, in order to be complete.
Then, we step to an analysis in terms of symbolic dynamics in that we analyze the series of liquid transported left
and right, i.e., we reduce flow to the fluid exchange
between the two rolls. This way we have a local measure by the stretching
and a more global one by the transport across the separatrix.

\subsection{Data Analysis}
\label{sec:dataanalysis}
\begin{figure}
\centerline{
\includegraphics[width=0.4\textwidth]{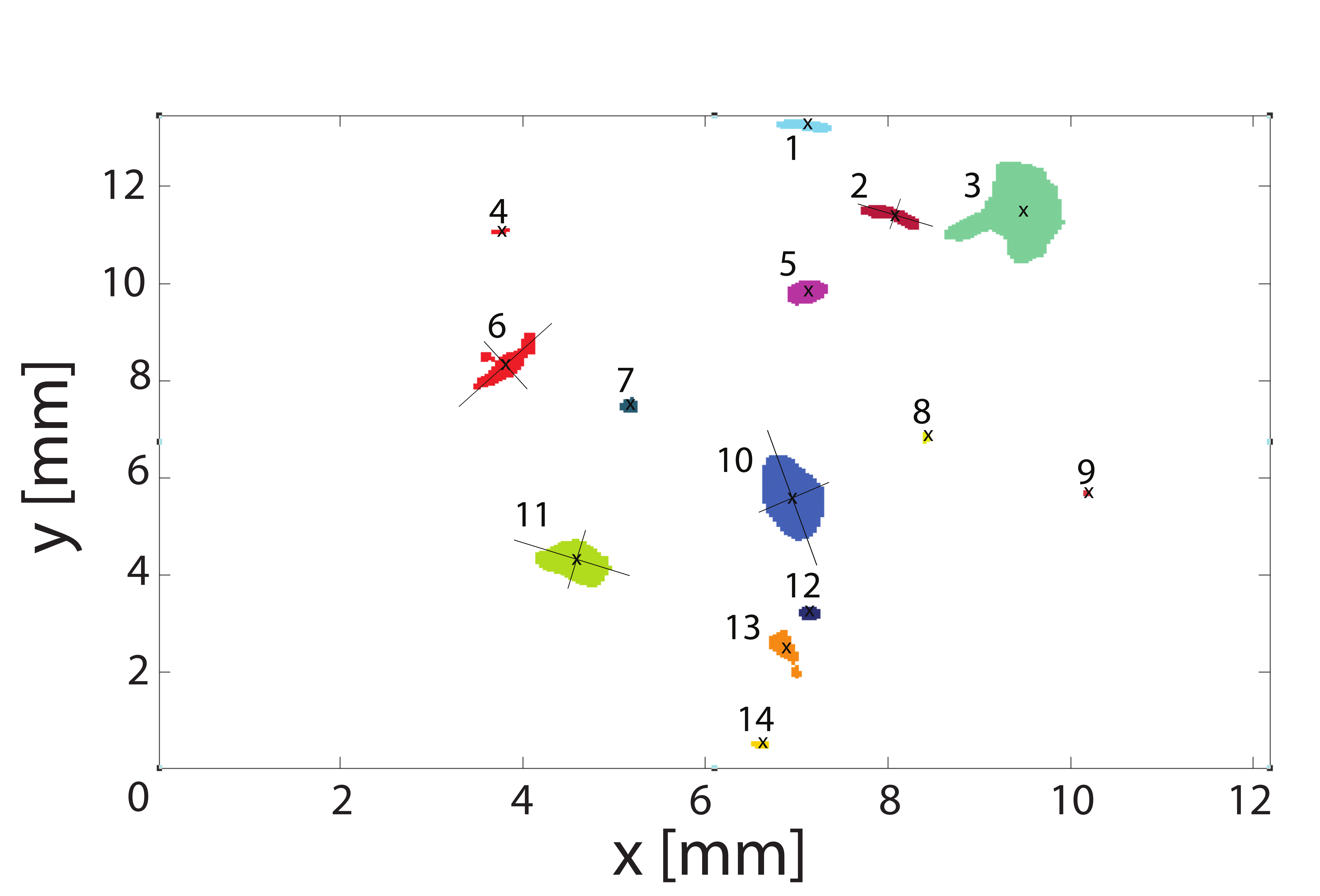}
}
\caption{example of the cluster identification for one frame, including center
of mass and selected principle components (not to scale).}
\label{fig:cluster_num}
\end{figure}

The captured video data is post-processed to enhance colors and contrast. To
analyze the behaviour of domains of the same thickness the video is converted
into a binary image where white corresponds to a single color respectively
thickness. As the spectrum repeats continually this technique is only valid if
the overall thickness deviation is smaller than a full period of the smallest
wavelength. Subsequently in each frame all clusters of the same thickness are
numbered and consecutively linked through the following frames (Fig.
\ref{fig:cluster_num}). This enables us to track the volume, velocity,
deformation rate and angular velocity of the moving fluid. For each cluster the
velocity is calculated using the shift of its center of mass per frame. The
deformation rate is calculated by determining the change of the scale of the
principal components per frame. Similarly the angular velocity is given by the
rotation of the principal component. Averaging over all frames then delivers
the
spatial characteristics of the flow field generated by the cooling tip. The
cluster finding algorithm is operating with a linear backwards memory of
variable depth. However for now a tracking of one frame backwards is sufficient
to maintain connectivity of each cluster through all frames. The memory needs
to be limited as merging or dividing clusters would be considered connected,
thereby distorting the velocity and deformation rate.

\begin{figure}
\centerline{\includegraphics[width=0.4\textwidth]{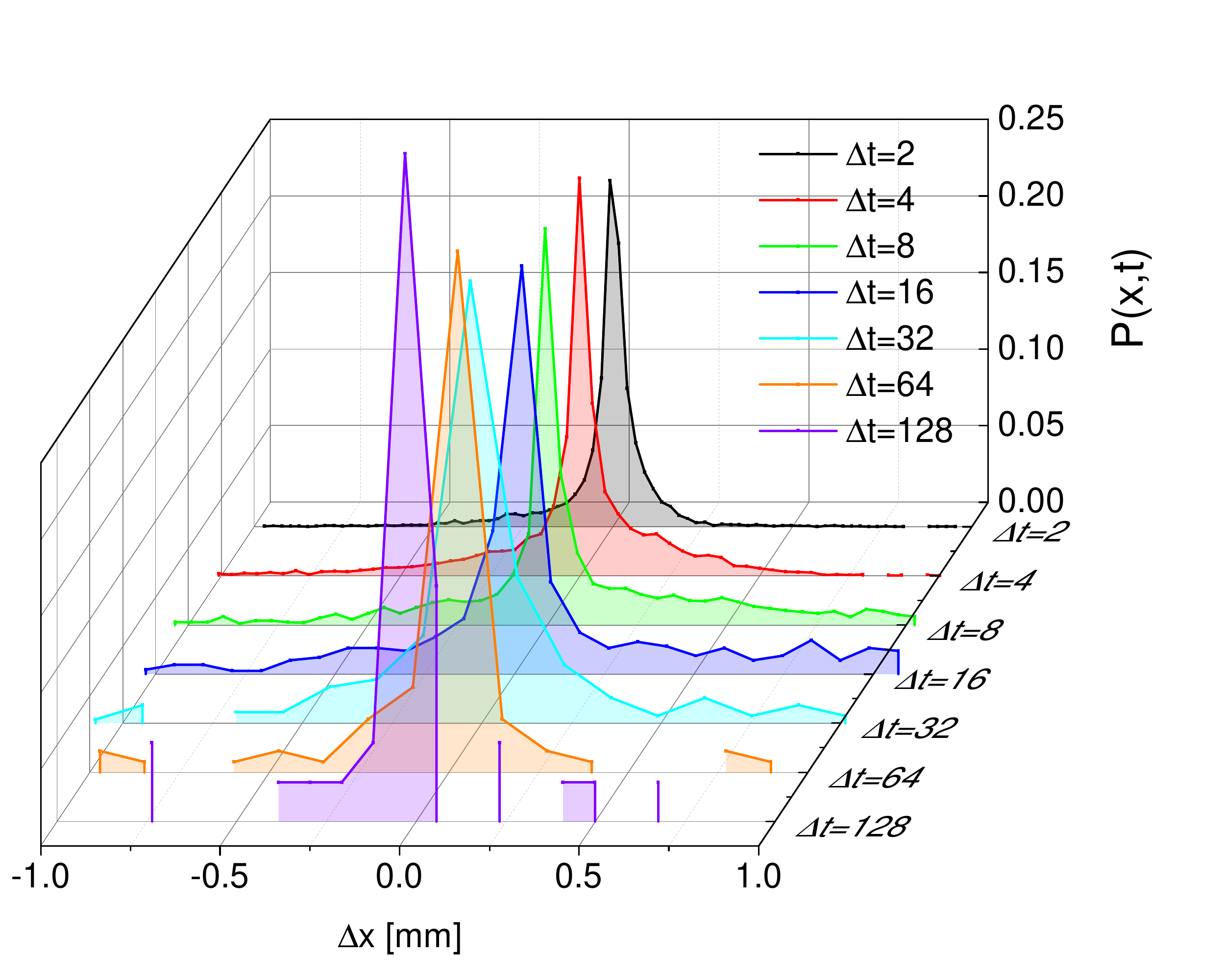}}
\caption{Probability $P(\Delta x,\Delta t)$ for different time scales $\Delta
t$(unit: frames, framerate: 100 frames per second). The transport is ballistic over one decade. Larger times are not shown due
to the insufficient number of data.}
\label{fig:pdxdt}
\end{figure}

The intermittency of the mixing film is captured by the probability $P(\Delta
x, \Delta t)$: we calculate the distance
$\Delta x$ for fixed $\Delta t$ (on logarithmic scale). The result is plotted in
Fig.~\ref{fig:pdxdt}. We observe basically ballistic transport for small $\Delta
t$, as could be expected because the more chaotic small scales are not resolved,
and the large scale transitions are not found by our cluster analysis.

\begin{figure}%
    \centering
        \begin{subfigure}{0.49\textwidth}
                \includegraphics[width=\textwidth]{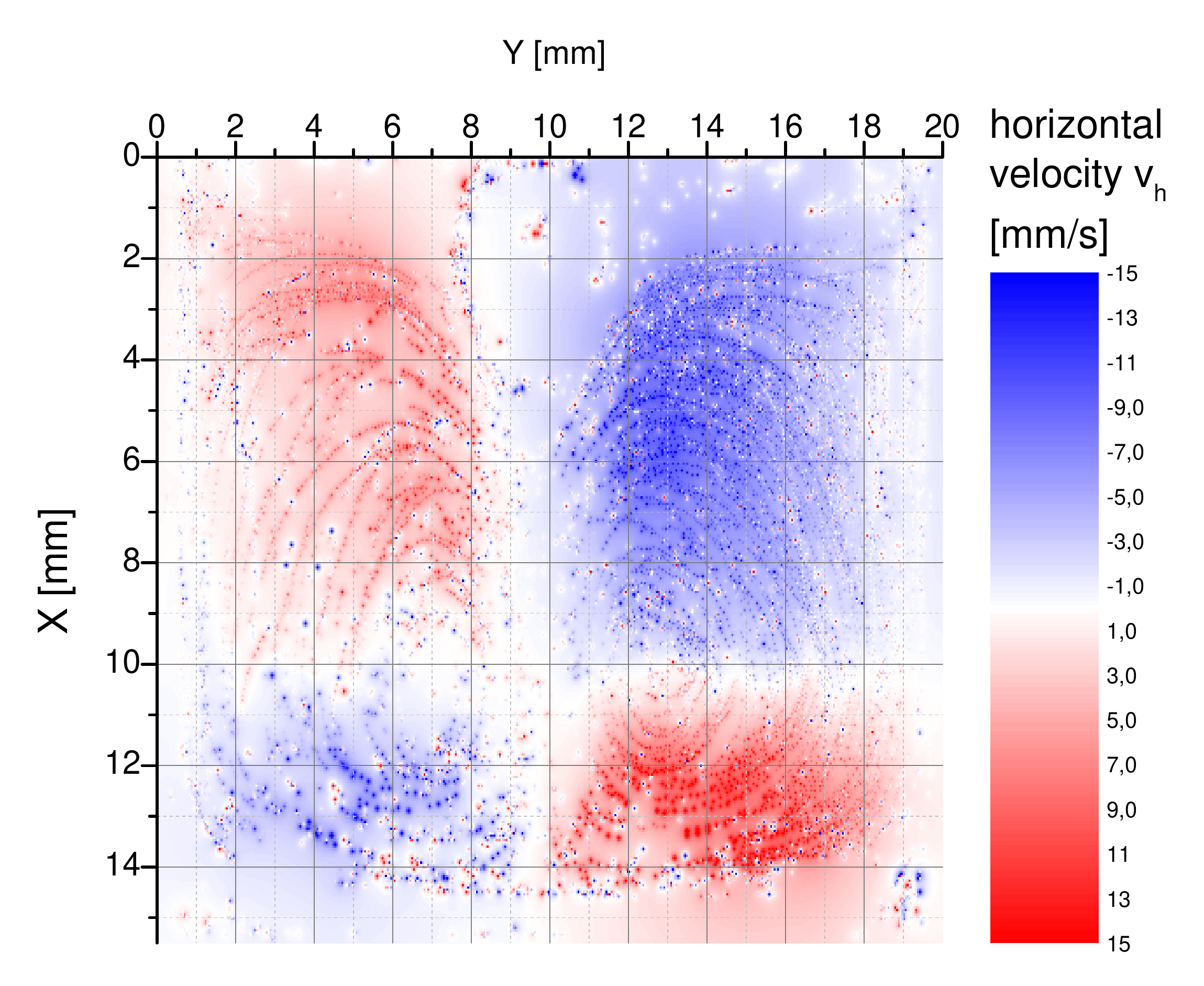}
                \caption{horizontal velocity}
        \end{subfigure}
        \begin{subfigure}{0.49\textwidth}
            \includegraphics[width=\textwidth]{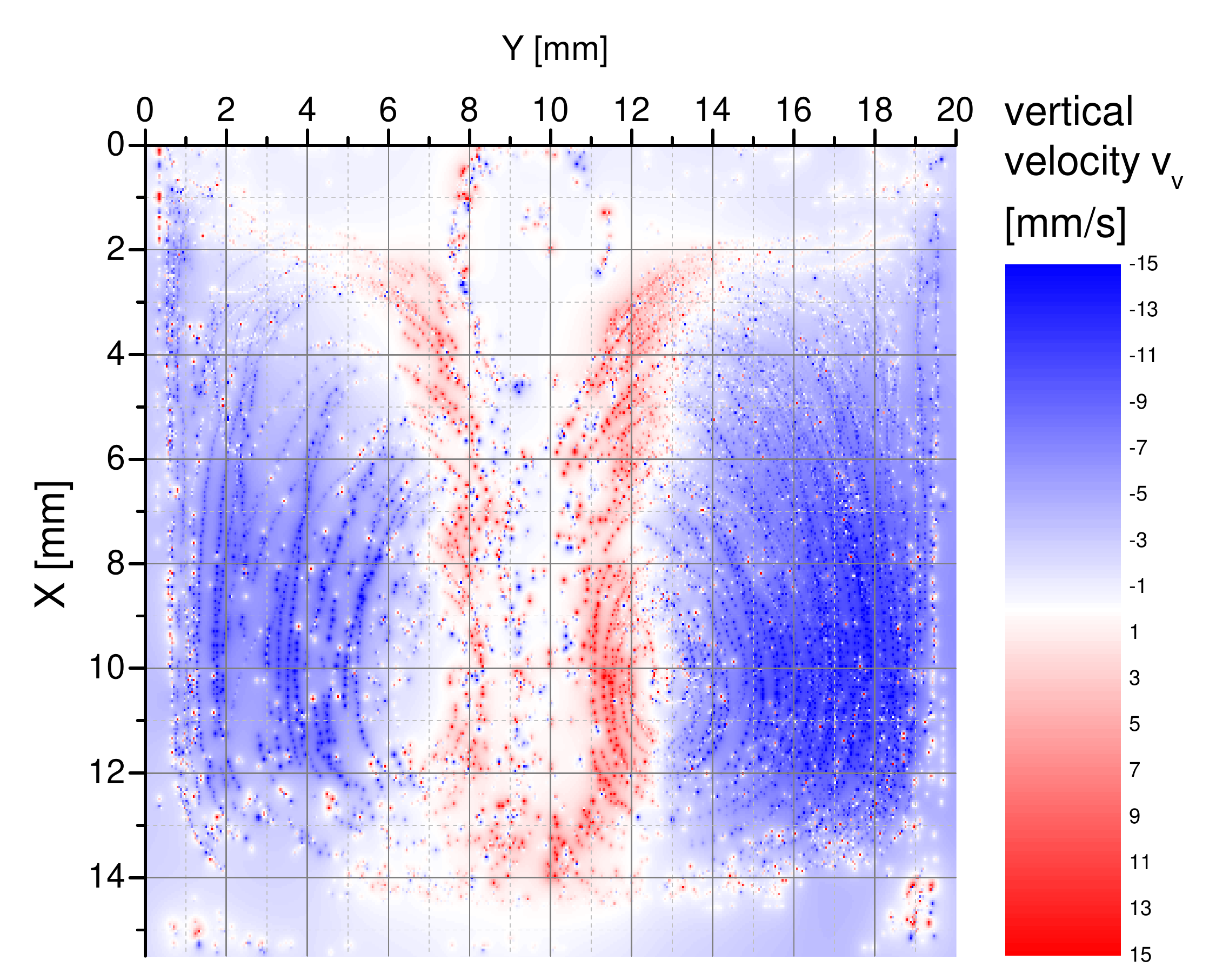}
            \caption{vertical velocity}
        \end{subfigure}
    \caption{averaged velocity field of the forced advection. The cooling rod is positioned at approximately $p_{rod}(x,y)=[2\,mm,10\,mm].$}%
    \label{fig:avg_velo}%
\end{figure}

In the generated velocity field (Fig. \ref{fig:avg_velo}) the fluid is
stretched
in regions of high velocity and compressed when it enters areas of lower
velocity. Due to the shear which is present between layers of different
velocity
folding happens. These two processes are the main aspects of mixing in a two
dimensional fluid. Diffusion processes can be neglected as the Reynolds number
is of the order of $10^3$.

The overall mixing can be characterized within each vortex as an averaging of
the stretching rate over the vortex area. The global mixing between the
vortices
is characterized by the probability of a fluid element to cross the separatrix.
This procedure can be seen in the view of effective diffusion, because the
averaged turbulent velocity field is a kind of random walker with possibly
anomalous and space dependent transition rates, which when averaged yield the
diffusion coefficient. The transition from one cell to the other is the minimal
setup, as discussed in
\cite{abel2000exit}.

We use a naive approach to analyze our film: take a spot of material of size
$\Delta x$, compute its time derivative approximatively from the time evolution
as finite difference and use then averaging to determine the macroscopic
properties. This is in contrast to studies with particles, where the velocity
of
relative dispersion is calculated, and only possible because we have already a
field at hand. From the clusters of the same thickness, we obtain the
eigenvalues $e_{1,2}^i$ along the principal axes; they correspond directly to
the size of the spot numbered $i$. Averaging the $e^i$ over time and filtering
yields an estimate for the fields $e_{1,2}(x,y)$. Now we compare that with the
definition of the diffusion coefficient:
\[
 D = \lim_{\Delta t\to 0} \frac{\left<(x(t+\Delta t)-x(t)^2\right>}{\Delta t}\;.
\]
In our situation the limit cannot be reached, due to sampling and consequently
we should use methods like the FSLE
\cite{Aurell-Boffetta-Crisanti-Paladin-Vulpiani-96}, which are ongoing work.
Here we show the results for the finite-size spots with $\Delta t = \Delta_s$
fixed to the minimal sampling time. We make of course an error in mixing
different spot sizes, which we counteract by choosing homogeneous spots of
similar size.

The space--dependent diffusion is then  estimated by
\[
 D_{est}= {\frac{\left<x(t+\Delta_s t)-x(t)^2\right>}{\Delta_s}}
\]
where averaging over many times ($\sim 10^6$) is performed. Of course this is a very rough
approach, but we will see that in a certain time range we obtain reasonable results,
and can compare this with the enhancement of thinning. This is, however not the full story: for
mixing it can be desirable to have faster than normal diffusion,
the characterization of a real process involves {\em finite} intervals $\Delta
t$, $\Delta x$, and the degree of anomaly is given by the scaling $\Delta t
\sim \Delta x^\alpha$, with $\alpha=2$ for normal diffusion. since we prescribe
$\Delta t$ and determine $\Delta x$ accordingly, the full statistical
characterization is given by the probability function $P(\Delta x,\Delta t)$, cf. Fig. \ref{fig:pdxdt}.

So, we characterize mixing within the two rolls by an effective diffusion
approach and the mixing between the two rolls by statistics of the crossings of
the separatrix in the middle of the
setup.

The fluid itself offers no contrast to track the motion of filaments. However,
the above described reflection imaging transforms the relative thickness
difference of convected filaments into a thickness map whose evolution can be
followed and analyzed. Compared to other techniques like seeding beads or
injecting ink, which are unpractical due to the resulting perturbation of the
thin liquid layer, the thickness tracking is non invasive and precise. A drawback
(not important here)
is that only a relative instead of an absolute thickness map is available and fine layering of
filaments is subject to diffusion, thus limiting the resolution.

In this study we investigate as well on the center stream in between the two stable vortices
which is deflected either to the right or to the left at the bottom of the
frame. At the cooling rod fluid from both vortices is formed into a center
stream. As the convection is bound by the frame the center jet is deflected
sequentially to the left or to the right at the bottom hyperbolic point. This
gives a relative measure of the amount of fluid mixed m,acroscopically between the left and
right  eddy. The binary series of transports can then be used to estimate the
efficiency of this mechanism by calculating the entropy and stochastic
properties.

\section{Methods and Results}
\label{sec:method}

In this section, we present methods and results based on the data obtained as described above.
A first subsection treats Lyapunov exponents, estimated by the stretching and folding of
small fluid elements. In this sense we only find finite-size values, which characterize the small-scale
properties of the flow (and not the infinitesimal ones, as one would need for the true Lyapunov-exponents).
Within that subsection we briefly recall and comment on the relation of measures from mixing theory and
dynamical systems.
In the second subsection results on the estimation of entropies are shown. We estimate the characteristic entropies
by the symbolic dynamics approach \cite{abel2000exit,kantz2004nonlinear} forming words of a certain length, compute their frequencies as estimators of
the probabilities, and eventually obtain an estimate for the entropy production with increasing word length.
At the end of this section we compare both measures using known relations.

\subsection{Lyapunov Exponents}
\label{sec:LE}

In dynamical systems theory, Lyapunov exponents (LE) are key quantities for the characterization of systems.
an equivalent is the efficiency in mixing theory for fluids. Essentially, in a fluid, the stretch of a fluid filament
is important, analogous to the stretching of a phase space element in a dynamical system.
In our case, both quantities are identical. Furthermore, the mixing efficiency is nothing but a normalized
version of the Lyapunov exponent for an ergodic system, since time and ensemble average need to coincide
for the equivalence to hold.

\begin{figure}%
\centering
\includegraphics[draft=false,width=\textwidth]{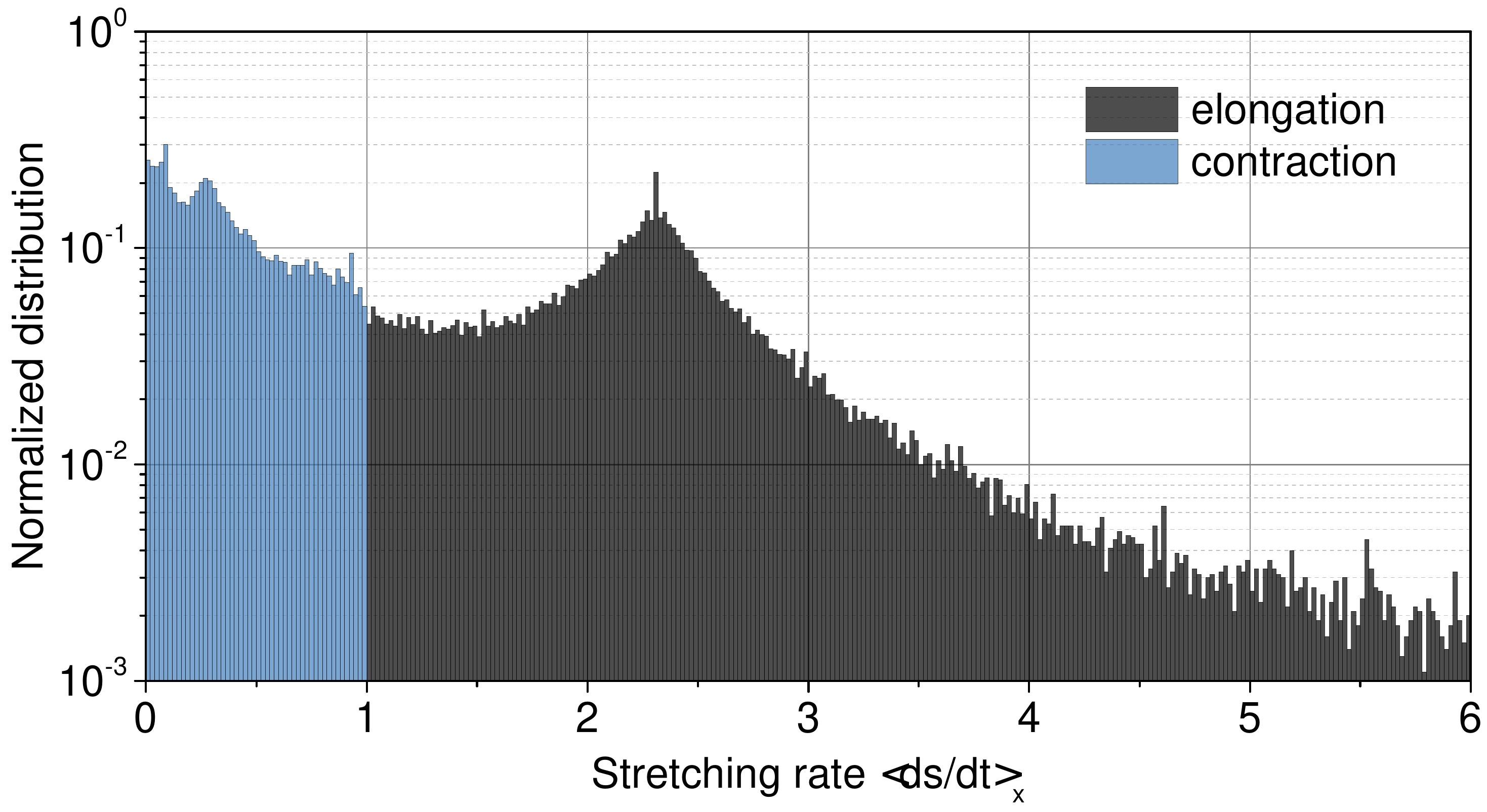}%
\caption[Stretching rate distribution]{Semi-logarithmic plot of the spatially-averaged stretching distribution.
Larger and smaller unity is marked differently for better visibility. The main difference lies in the behaviour for
contracting values, which can be zero, i.e. the spot observed vanishes. This is reminiscent of our data analysis,
explained above and is not expected for  a perfect measurement. }%
\label{fig:stretch}%
\end{figure}

In the following we describe how to estimate mixing efficiency
and LEs $\lambda$ starting from the stretch $s$ of a fluid filament along its trajectory, where
\begin{equation}
s = \lim_{\Delta x_0 \to 0} \frac{\Delta x_t}{\Delta x_0}\;
\label{eq:length_stretch}
\end{equation}
for the fluid velocity measured in the Eulerian (i.e. laboratory) frame. $\Delta x_0$ denotes the initial separation of two ``fluid points`` at time $t=0$, and $\Delta x_t$  the corresponding separation at time $t$. To satisfy the limit towards elements $\Delta x_0$ of infinitesimal length, the cluster data was filtered to contain only the smallest clusters available.
The local stretching rate $\frac{ds}{dt}$ (cf. Fig.\ref{fig:stretch}) is measured in a moving reference system (Lagrangian), therefore the velocity of the separating edges of a fluid filament are observed.

\begin{figure}
\centering
\includegraphics[draft=false,width=\textwidth]{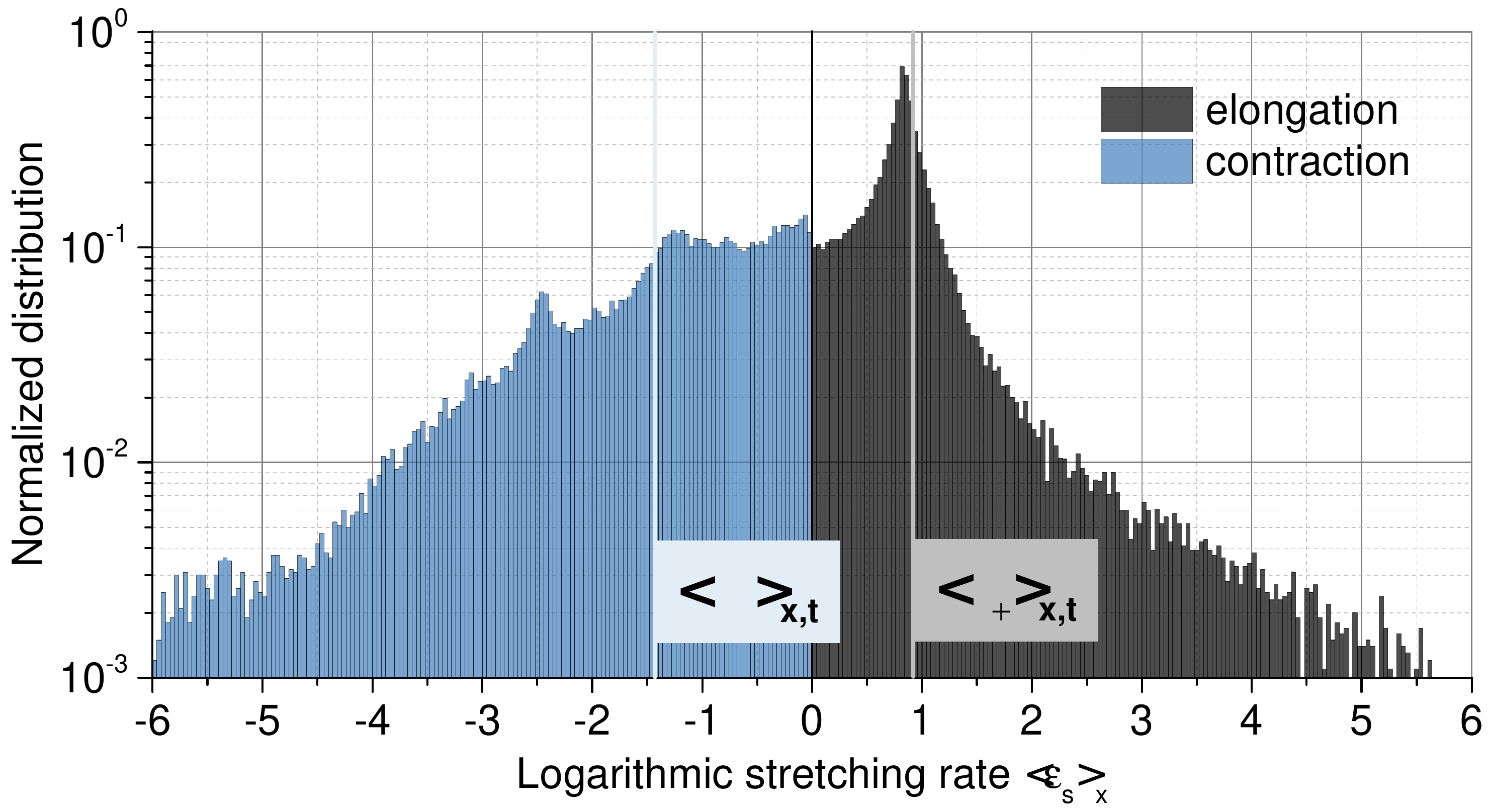}%
\caption[LE]{Distribution of the stretching rate $\epsilon_{s}$. The average over time gives the Lyapunov exponents $\left< \lambda_{-} \right>_{x,t}=-1.43 \; ,  \left< \lambda_{+} \right>_{x,t}=0.92 $, indicated by the vertical lines. Clearly, the distribution of negative and positive LE ist different.}%
\label{fig:specific_stretch}%
\end{figure}

The Lyapunov exponents $\lambda$ describe the typical property of chaotic systems, in particular 2D systems with hyperbolic points: the trajectories of two nearby points diverge exponentially. We can relate that to fluid motion, for details cf. e.g. \cite{ott2002chaos,tel2006chaotic}.
The exponential divergence for infinitesimally small times is expressed as $\Delta x_t\sim {\Delta x_0} e^{ \lambda t}\;.$
The local Lyapunov exponent is then given as the long time average of the logarithmic stretching rate (cf. Fig. \ref{fig:specific_stretch}):
\begin{equation}\label{eq:LE}
  \epsilon_{s}=\frac{D \ln s}{Dt} \; \rightarrow \;
  \lambda_l = \lim_{t\rightarrow\infty} \frac{1}{t} \int\limits_0^t \epsilon_s \mbox{d}t'\;,
\end{equation}
where we recognize the difference to the stretch (eq.\ref{eq:length_stretch}) in averaging the logarithm or the ratio of the lengths directly.
For ergodic systems, time and (phase) space average are identical and we can compute alternatively
\begin{equation}
\left< \lambda_{} \right>_{x,t}= \left < \lim_{\Delta x_0 \to 0} \lambda_l \right>_x \;.
\label{eq:LE_avg}
\end{equation}
 with the notation $\left <\, \right >_x$ for the space average. We can compute the two Lyapunov exponents from the contraction (-) and elongation (+) of the tracked fluid filaments and obtain as an estimate on the overall mixing properties of the flow:
\begin{equation}\label{eq:LE_averages}
  \left< \lambda_{-} \right>_{x,t}=-1.43 \; ,  \left< \lambda_{+} \right>_{x,t}=0.92 \;.
\end{equation}
If the flow is conservative $\||\lambda_{-}/\lambda_+ =1\||$, here we find a contraction considerably
smaller than $-1/0.92 \simeq -1.087$.  It is explained by the fact that we do not only measure the stretch, but in addition the general volume loss
of a spot of a certain color due to film thinning and diffusion of the thickness map, as explained above, cf. Fig.\ref{fig:cluster_num}; this contributes to the negative LE.

The distribution of the positive and negative LE are quite different. This is explained by the dynamics: elongation -positive LE- takes place
mainly in the rapid flow region between the two fixed points (the separatrix). This leads to a very pointed distribution with faster than exponential decay, cf. Fig.~\ref{fig:specific_stretch}, black marked part. In contrast to this contraction happens over the whole region of the film
with quite different and much slower dynamics. Additionally, the contraction rate data has a lower signal to noise ratio compared to the elongation rate, as the absolute values are smaller and closer to the resolution limit of the setup.

Before stepping to the estimation of the entropy we want to describe the connection between LE and mixing efficiency,
both quantities are closely related:
 \begin{equation}\label{eq:stretch_eff}
    \left<E_s \right> = \frac{1}{t} \int\limits_0^t \frac{\epsilon_s}{\sqrt{\vec{D}:\vec{D}}}\mbox{d}t'
\end{equation}
We see that the difference lies in the nhe normalization. It involves the
symmetric part of the velocity gradient $\vec{D} =  \frac{1}{2}\left [ \nabla
\vec{v}+(\nabla \vec{v})^T \right ]$ (equivalent to the stretching tensor),
where $\vec{D}:\vec{D}$ needs to be constant over the pathlines, for this
expression to be valid. It is used as a normalization to obtain a stretching
efficiency smaller or equal to one. Using this normalization, one can
demonstrate \cite{ottino1989kinematics} that the upper bound of the efficiency
for two-dimensional flows is $E_{s ,max}=\sqrt{2}/2=0.707$
\cite{ottino1989kinematics}. Since the stretching tensor is not readily
available from the cluster data and is only used as a normalization in Eq.~\ref{eq:stretch_eff}
it was not used here and we refer to the unnormalized
Lyapunov exponents.

\subsection{Entropy }
\label{sec:entropy}

Similar to the Lyapunov exponent the entropy production of a dynamical system is a measure of its chaoticity. Here we compute the Kolmogorov-Sinai entropy $h_{KS}$for the center jet deflection and compare it to $\left< \lambda_{-/+} \right>_{x,t}$ which we obtain from the stretching rates of the fluid filaments. The topological entropy $h$ is the upper limit for average Lyapunov exponent: $h \geq \left< \lambda \right>$ \cite{tel2006chaotic} and the Pesin formula equates the topological entropy with the sum of all positive Lyapunov exponents: $h=\sum_{i=1}^n \lambda_{+,i} > h_{KS}$, and sets the limit for $h_{KS}$ \cite{ruelle1989chaotic}. More precisely the relation between the topological entropy $h$ and $h_{KS}$ is given via the Variational Principle: $h(T)=\mbox{sup}\left(h(\mu)_{KS}: \mu \in \mathcal{P}_T(X) \right)$, where $\mu$ ranges over all T-invariant Borel probability measures on X. Thus $h$ is an upper limit for the KS-entropy.
A direct relation between Lyapunov exponents and entropy is available via the information dimension
 \begin{equation}\label{eq:young}
        D_I=h(\mu)\left(\frac{1}{\left< \lambda_{+} \right>_{x,t}}+\frac{1}{|\left< \lambda_{-} \right>_{x,t}|}\right)
\end{equation}
 of an ergodic invariant measure $\mu$ of a smooth invertible map with Lyapunov exponents $\left< \lambda_{+} \right>_{x,t} > 0 > \left< \lambda_{-} \right>_{x,t}$ \cite{young1982dimension,ott2002chaos}.  $\left< \lambda_{-/+} \right>_{x,t}$:  Lyapunov exponents calculated from stretching (+) and contraction (-) rates.

Comparing Eq~\ref{eq:young} to the Kaplan-Yorke conjecture \cite{frederickson1983liapunov}:  $D_L = 1 + \left< \lambda_{+} \right>_{x,t}/|\left< \lambda_{-} \right>_{x,t}|$ , with Lyapunov dimension $D_L$, suggests that $h(\mu)=\left< \lambda_{+} \right>_{x,t}$ in the case of the natural measure of a 2D smooth invertible map.

Let us now compute the values for the entropy to the degree our data allow with respect to length of the time series and accuracy.
First, we need to define a kind of ``alphabet'' to identify words. Our procedure follows essentially \cite{abel2000exit} and
for more details we refer to that publication.
The jet is deflected either into the left or into the right vortex at the bottom of the frame. We analyze the deflection pattern to check for deterministic components, which would lower the mixing efficiency of this fluid transport.

The sequence of left (treated as 0) and right (treated as 1) transports is a binary time series from which one can associate a word of length $n$, out of a finite alphabet:  $W_k^n=(S_k,S_{k+1},...,S_{k+n-1})$.
The block entropies $H_n$ are then calculated from the word probability distributions $P(W^n)$:
\begin{equation}\label{eq:block_entropy}
  H_n = - \sum_{W^n} P(W^n)\ln\,P(W^n)
\end{equation}
where $W^n$ represents the set of all possible words of length n. The entropy
per unit time is defined as:
\begin{equation}\label{eq:entropy_time}
\begin{split}
  h_n & =H_{n+1}-H_n \\
  h_{KS} &=\lim\limits_{n \rightarrow \infty}{h_n}
\end{split}
\end{equation}

\begin{figure}[ht]
 \centering
        \begin{subfigure}{.49\textwidth}
            \includegraphics[width=\textwidth]{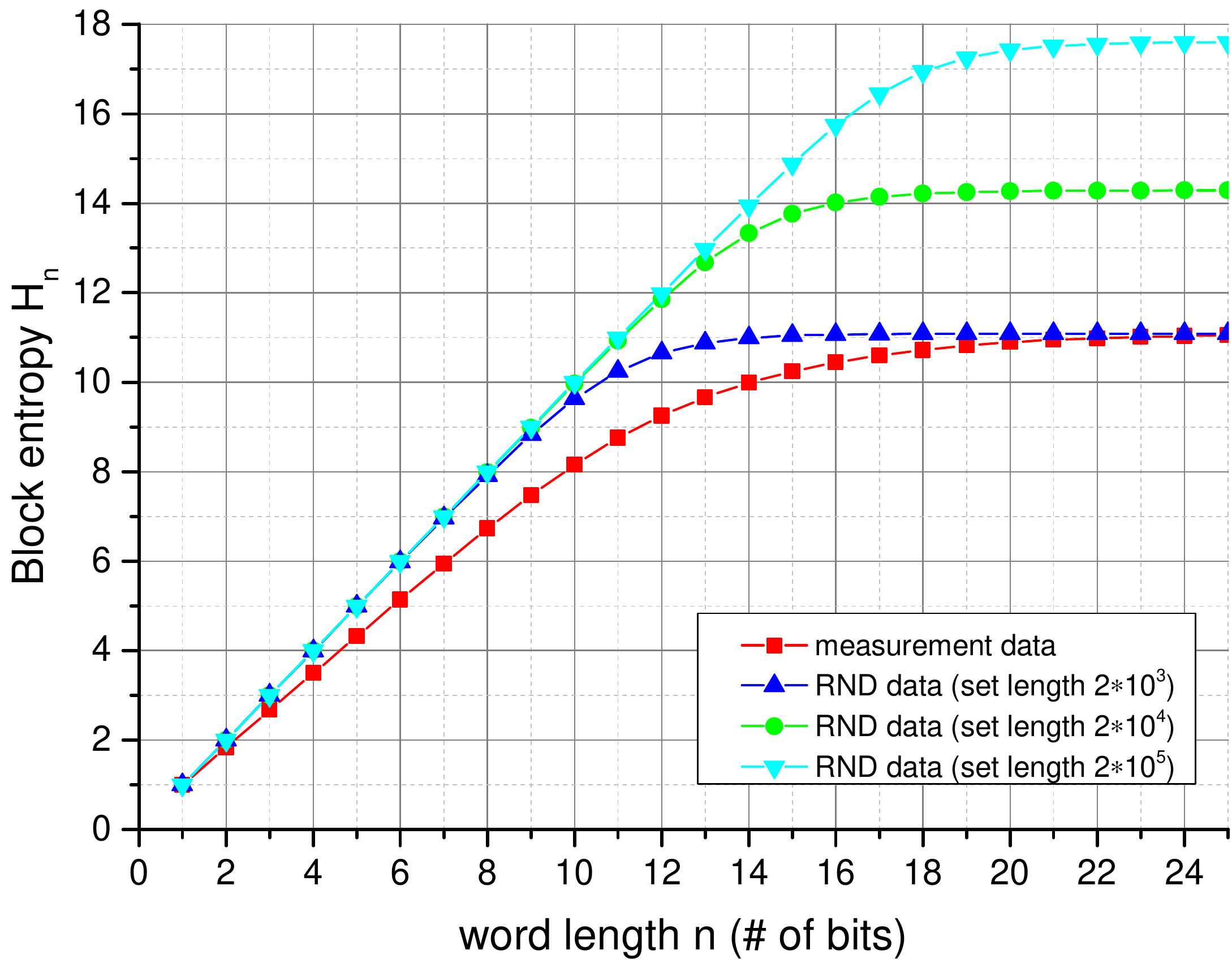}
            \caption{Block entropy $H_n$}\label{fig:entropy}
        \end{subfigure}
        \begin{subfigure}{.49\textwidth}
            \includegraphics[width=\textwidth]{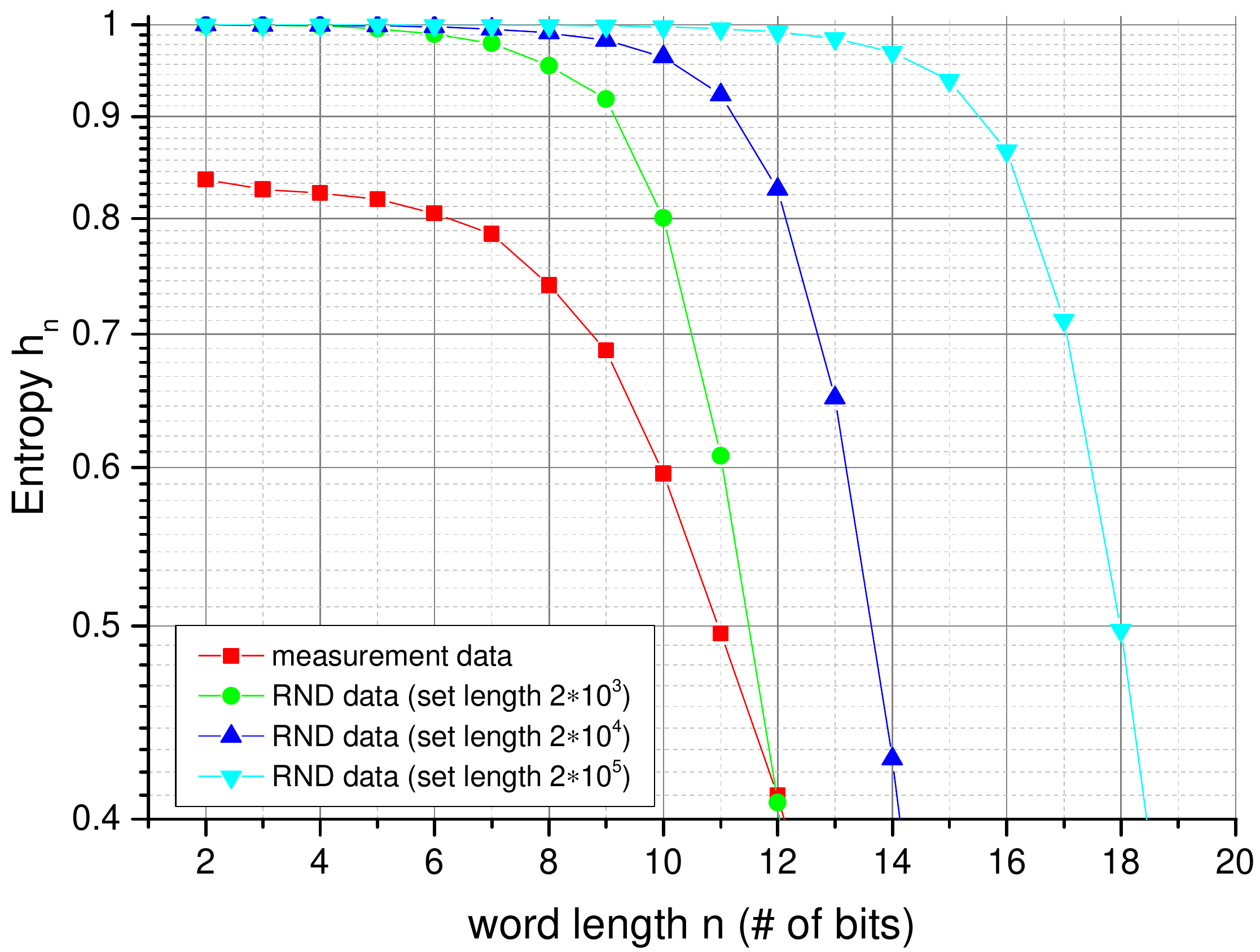}
          \caption{Entropy $h_n$}\label{fig:entropy_gain}
        \end{subfigure}
        \caption{Block entropy $H_n$ and entropy $h_n$ with increasing word length of the measurement set (set length $2 \cdot 10^3$), compared to random distributions of variable set length.}
\end{figure}

In the presented  case the practical limit of $h_{KS}$ is given by the finite series of events thus truncating the number of possible words and the possible entropy gain by increasing the word length.
$h_n$ can be interpreted as the rate of information production, which for a finite signal decreases when the combined length of all possible words $W^n$ becomes larger than the signal itself: $\sum_{W^n} n > l_S$.  The mixing efficiency is maximal when the jet deflection is equally distributed for left and right. Therefore, we use a set of equal left/right probability as a benchmark for the measurement data . As expected, the entropy $h_n$ remains constant with increasing word length until the finite size limit of the data set is reached, and longer sets hit this limit at higher word lengths, cf. Fig. \ref{fig:entropy}.

 The entropy $h_n$ of the different distributions below this limit is identical, in particular for the data with the same length as the measurement set. Therefore, the deviation in $h_n$ of the measurement data when compared to the equally dstributed data sets cannot simply be omitted as a finite size effect. $h_n$ of the measurement data is approximately 15-20\% below the possible maximum given by the random sequences, which indicates a non-uniform process or deterministic behaviour, cf. Fig. \ref{fig:entropy_gain}.

\begin{figure}[ht]
  \centering
    \begin{subfigure}{.49\textwidth}
      \includegraphics[width=\textwidth]{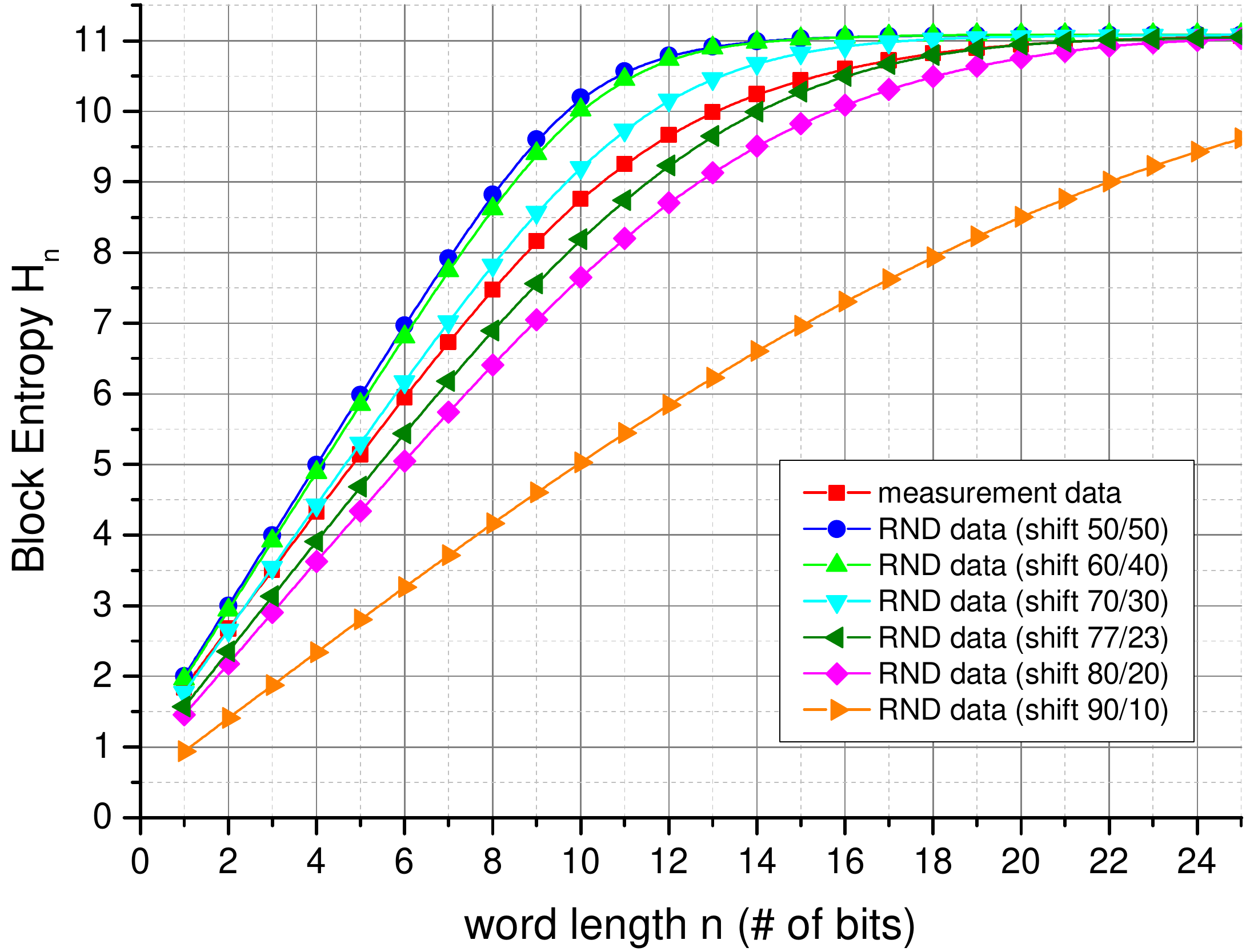}
      \caption{Block entropy $H_n$}\label{fig:rnd_skewness}
    \end{subfigure}
    \begin{subfigure}{.49\textwidth}
      \includegraphics[width=\textwidth]{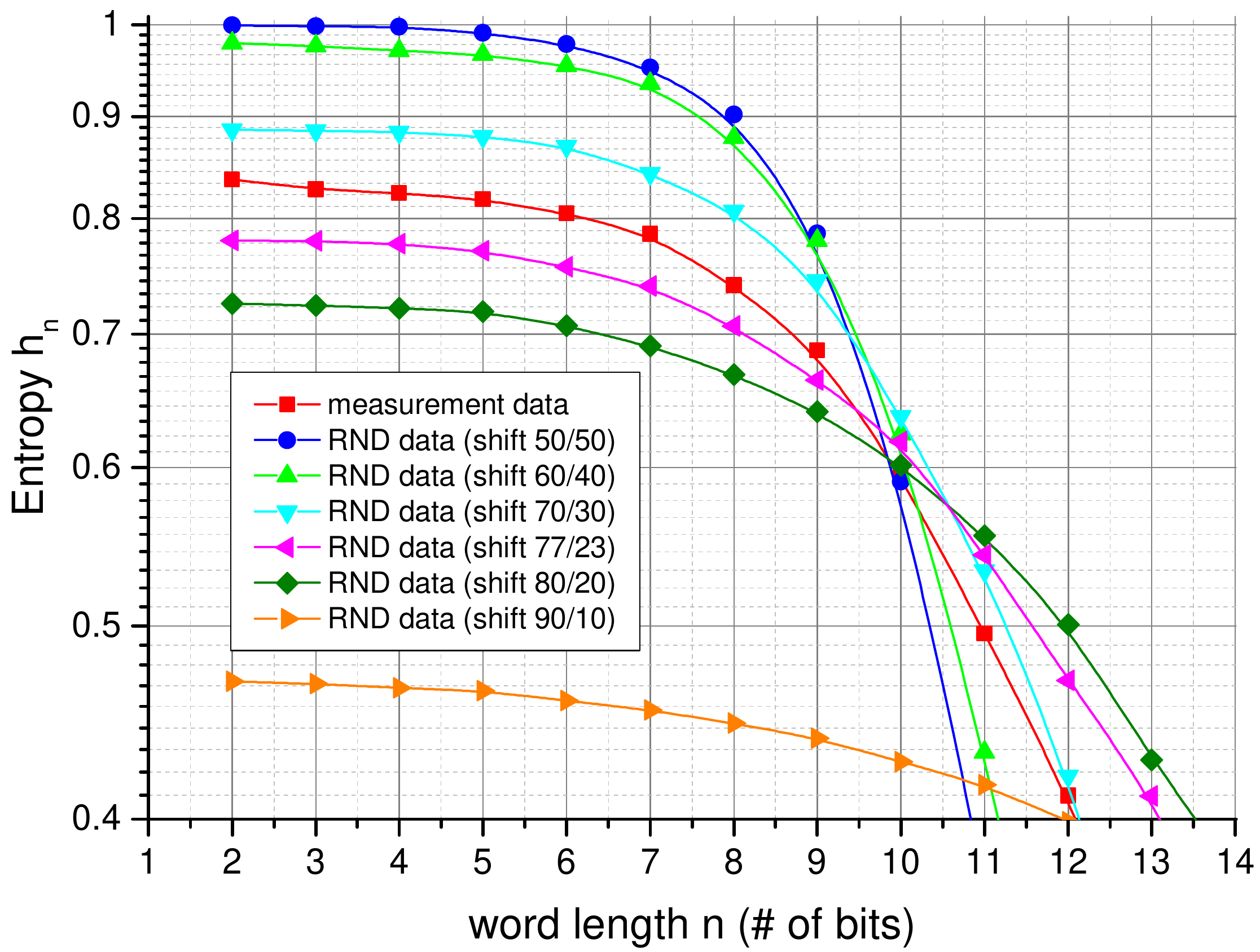}
      \caption{Entropy $h_n$}\label{fig:entropy_gain_shift}
    \end{subfigure}
    \caption{Block entropy $H_n$ and entropy $h_n$ of the measurement data compared to random distributions with varying skewness. Data set length is constant.}
    \label{fig:entropy_shift}
\end{figure}

One might assume that the sequence is tilted towards one side and that this preference causes the flow pattern to be more predictive. To check for this property we compared the data set to skewed random distributions which have a lower entropy production due to the higher predictability of the signal. In Fig.~\ref{fig:entropy_shift} random sets with an uneven distribution up to a probability of $P=0.9$ for one direction of the flow are compared.
The entropy production of a $77/23$-skewed distribution is comparable to the measurement data. In this case the deflection of the center jet would be directed towards one side approximately 8 out of 10 times. However, the distribution of left and right transports in the measurement data is even which shows that the deterministic components of the signal is not due to a simple asymmetry in the experimental setup.

\paragraph*{conditional probabilities}

\begin{figure}[ht]
          \begin{subfigure}{0.49\textwidth}
                \includegraphics[width=\textwidth]{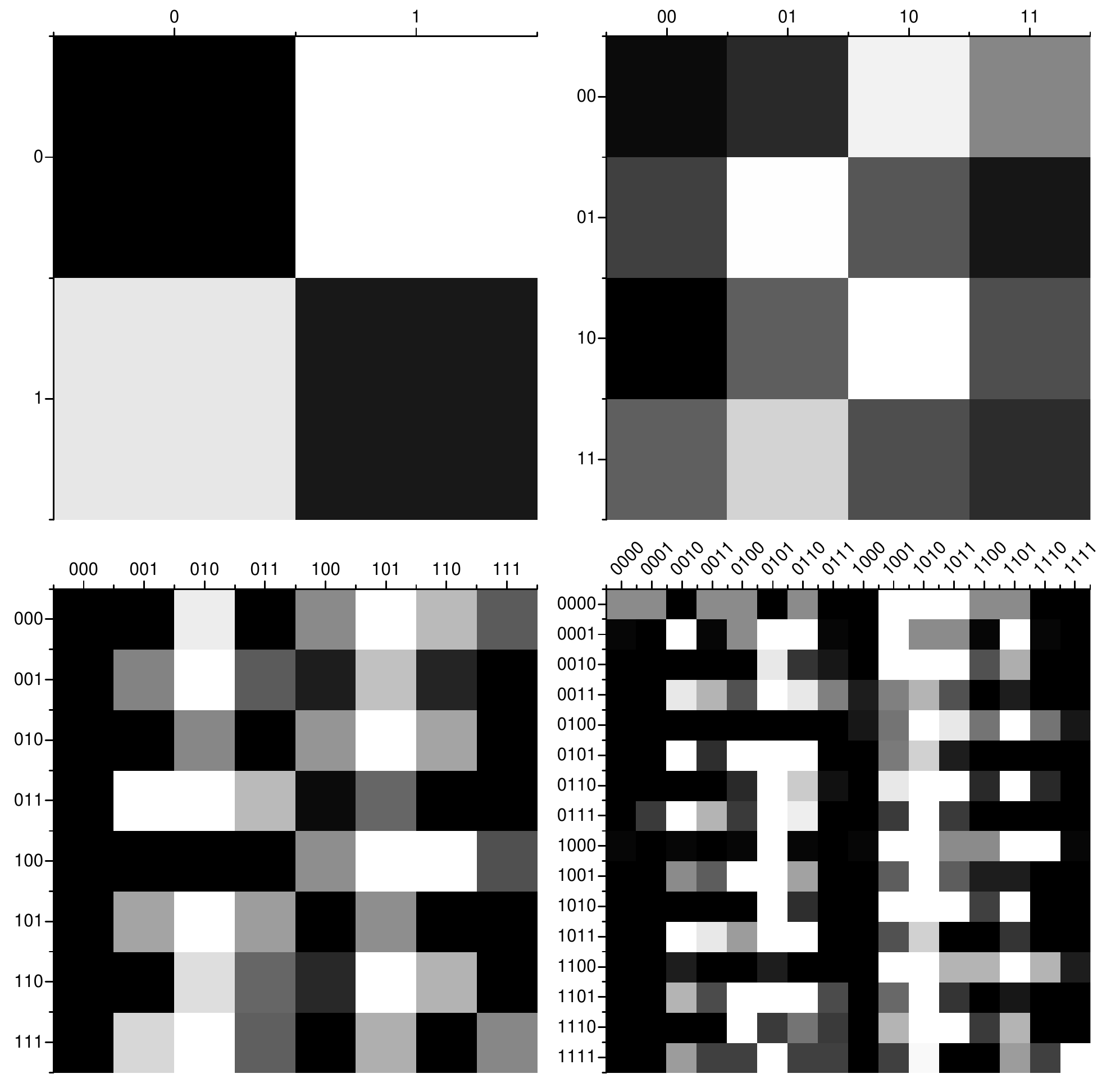}
                \caption{Measurement data}
                \label{fig:conditional_prob_data}
        \end{subfigure}
        \begin{subfigure}{0.49\textwidth}
              \includegraphics[width=\textwidth]{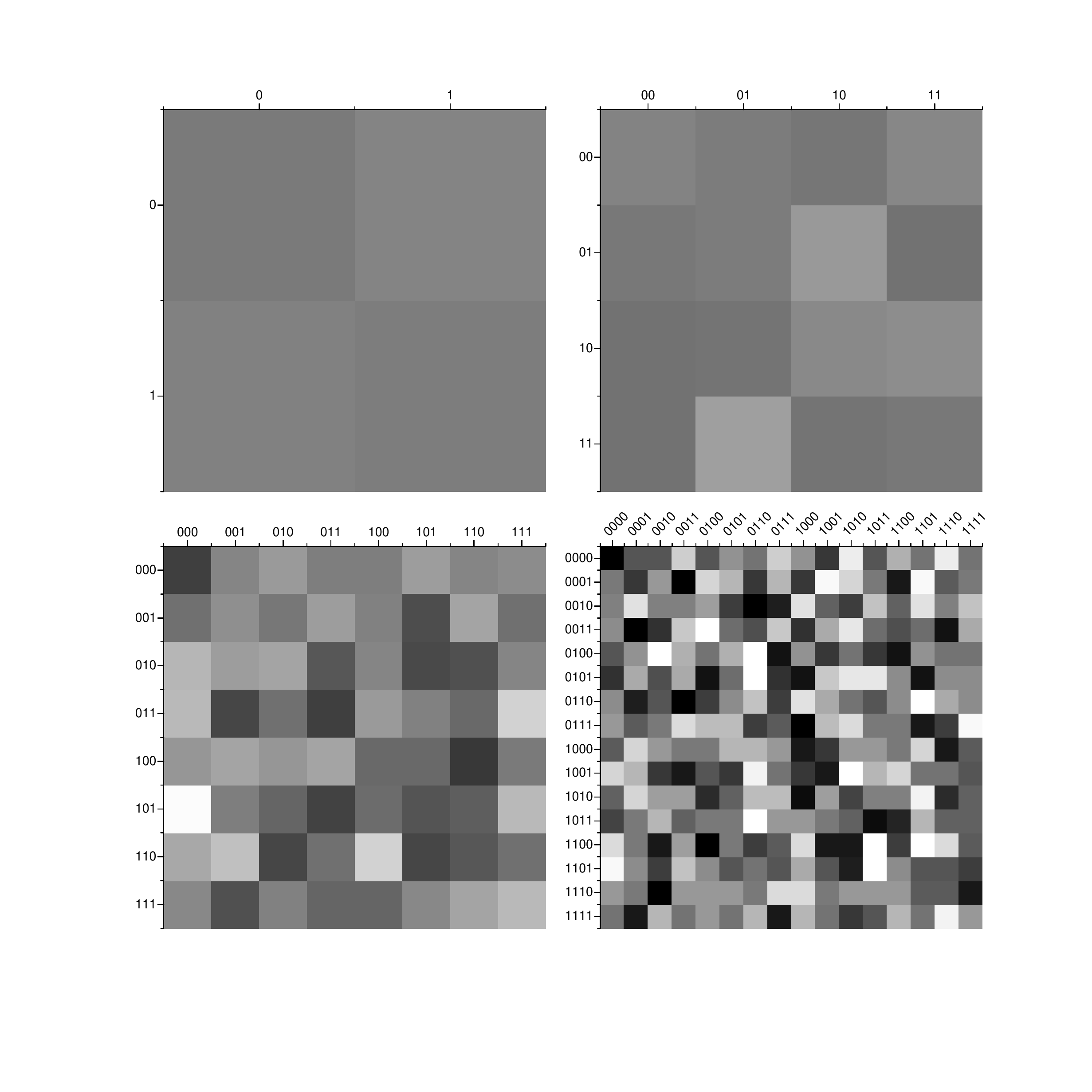}
              \caption{Random distribution}
              \label{fig:conditional_prob_RND}
       \end{subfigure}
       \caption{Conditional probabilities for measurement data (a) compared to a random distribution (b). Each tile represents the conditional probability $P(W^m|W^n)$, where $W^m$ is plotted on the x-axis and $W^n$ is plotted on the y-axis. Word length $l_W =1..4$. The grey scale for each plot is centered at the expected probability, i.e. for $l_W=2$: $P(W^m|W^n)=0.25$ for all possible combinations. From this value the scale covers $50\%$ deviation, i.e. $l_W=2$: black: $P(W^m|W^n)=0.125$, white: $P(W^m|W^n)=0.375$.}
    \label{fig:conditional_prob}
\end{figure}

Another option to check for deterministic components is to look for recurrent sequences in the signal.
Therefore, we calculated the conditional probabilities $P(W^n|W^m)$  with $n=m =1..4$ to look for recurring patterns with a memory of up to 4. This limit applies due to the finite data set size, therefore the total word length $l_W=n+m$ for our measurement is restricted to 8. For each word length the conditional probabilities of all possible combinations were calculated giving a matrix of size $n\,\times\,n$. Again we compare to a random signal of the same size as the measurement data set. For a random data set of infinite length one expects an even distribution of conditional probabilities. However, as we compare finite size data sets, the conditional probabilities become nonuniform at higher word lengths as not all combinations are represented equally, cf. Fig. \ref{fig:conditional_prob_RND}. The figure displays the deviation from a perfect random distribution, which would show as a 50\% gray tile.

The conditional probabilities of the data set deviate substantially from the even distribution the random data provides, cf. Fig.~\ref{fig:conditional_prob_data}. The jet is more likely to alternate between left and right transports which is evident by the higher conditional probabilities of alternating binary combinations, ie. $P(W^n|1010),P(W^n|0101) $. Some combinations with two or more consecutive transfers in the same direction show a higher than average probability, but these are always combined with alternating sequences. Uniform combinations, ie. $P(W^n|0000),P(W^n|1111) $, are extremely rare events.

The diminished entropy production with increasing word length is caused a preference for alternating transfers which are evidenced by nonuniform conditional probabilities. The signal is more predictable and  most common combinations are already covered with lower length words thus the additional words carry less new information.

Eventually, we compare the results from entropy and LE estimation. For the LE we obtained
$\lambda_+ = 0.92$, $\lambda_-=-1.43$. For the Shannon entropy we can read off the entropy gain from Fig.~\ref{fig:entropy_shift} (red line and markers) as $h_{KS}=0.82$. Using the Kaplan-Yorke conjecture we should find a coincidence $h\simeq \left<\lambda_+\right>$ with a Kaplan-Yorke dimension of 1.92. Given the error sources in the LE estimation and the relatively short time series, we consider this result as a very good coincidence.

\section{Discussion and summary}

We presented a relatively simple experiment exhibiting complex dynamics, where turbulent mixing reaches out for the nanoscale,
at least in one dimension. This renders the flow two--dimensional, however with additional forces
acting between the surfaces: disjoining pressure and capillary pressure.

With respect to turbulence, we observe that the relatively low-$Ra$ convection
generates a weakly turbulent flow with two prominent rolls. the turbulence mainly can be observed inside the rolls on shorter time scales, whereas transport between the rolls is on a slow scale and shows signs of chaos.
To characterize the flow, we take advantage of our
measurement capacities, and the color imaging velocimetry (CIV) technique.
In addition to the flow field, concluded from Lagrangian trajectories, we
obtain local deformation rates which are used to estimate the Lyapunov exponents.
With respect to mixing we clearly observe the typical chaotic filamentation.
As mentioned in the context of weak turbulence, global
mixing on the large scale happens between the left and right convection roll.

We focused on two different methods to quantify the dynamics of the flow:
Lyapunov exponents and entropies. Whereas the LE are calculated using small-scale dynamics
using CIV, i.e. tracking of small fluid spots, entropies have been calculated
using the left-right transport across the separatrix. Qualitatively, these two
observations are related by the two ``spots'' generating the dynamics:
the hyperbolic fixed points. They determine the microscopic properties (LE) and
as well the transport across the separatrix. Thus, based on dynamical systems principles
one expects that both quantities coincide. Given the two different approaches and
the very different spatial scales, this coincidence is a formidable cconfirmation
of theory.

The presented data is gathered from 8 individual runs of the experiment.
Entropy production and conditional probabilities were calculated for the data
set and compared to truly random data sets of varying size and skewness. The
conditional probabilities show that an alternating pattern of left and right
transports is preferred which increases the predictability and lowers  the
entropy production of the measurement data with respect to equally distributed
data. The number of left and right transports in the data is even with a
deviation of $2\%$ from the exact left/right randomly equal  distribution, which can be accredited
to inaccuracy of experimentally obtained data (rod not positioned at the exact
center) and limited data set length. Although the entropy production $h_n$ is comparable to skewed random distribution with approximately 70/30 shift, these are two unlinked deterministic mechanisms which lead to a decrease of $h_n$.

Due to the shortness of the time series, an entropic analysis has errors beond a certain
length of the words formed. In order to have a quantitative measure for this
length we compare the measurement with numerically determined left/right
random sequences without memory.
As displayed by the random data sets of varying size a larger data set would
yield a higher accuracy in the entropy production rate and conditional
probabilitiyes, but the fundamental scaling behaviour remains the same. Thus the
statistical evaluation of the measurement data is valid.

Using such kind of numerical validation procedure, we find clearly that the asymmetry found should be attributed to
higher-order memory in the data. This is quite plausible, since we are investigating
a fluid system, where memory is ``built in'' for advected particles and so for advected fluid
particles, too.
The calculated KS-entropy serves as an upper limit to the Lyapunov exponents,
which were calculated from the stretching rates. Experimentally, we used color-tracking
which focuses on small scales.
This area tracking algorithm provides detailed information about the flow field
and is adaptable to work on any data of deformable clusters with high enough
contrast with respect to their surrounding.

One key question regards the consistence of LE and entropy estimate. We found an astonishingly
close result for entropy and LE using Kaplan and Yorkes conjecture. Both
agreed within an error of 10\%. The values of 0.92 and -1.43 indicate a relatively
moderate large-scale mixing which is due to the fact that the setup involves only two
main flow regions among which one has to achieve mixing.

However, in practical terms, such nanofluidic devices can be well used as free-standing
mixers with very flexible surface properties. The mixing times are then quite fast, since
the device dimensions can be miniaturized further. The true advantage lies, in our opinion,
in the very good ratio of surface and volume of the film: such a film has an enormous surface,
in our setup  the ratio was 1:100000, which is probably only achievable with thin films. The free-standing property, finally,
allows to avoid problems with solid-liquid interface forces, rather one can focus on a chemically
favorite design of the equally treated surfaces.

Generally, we think that with our relatively simple experiment we can touch deep
physical questions on the microscopic nature of thin film flows and treat at the
same time very practice-oriented topics.

\bibliographystyle{plain}
\bibliography{all}

\begin{thebibliography}{10}

\bibitem{abel2000exit}
M~Abel, L~Biferale, M~Cencini, M~Falcioni, D~Vergni, and A~Vulpiani.
\newblock Exit-time approach to $\varepsilon$-entropy.
\newblock {\em Physical review letters}, 84(26):6002, 2000.

\bibitem{atkins2010investigating}
L.J. Atkins and R.C. Elliott.
\newblock Investigating thin film interference with a digital camera.
\newblock {\em American Journal of Physics}, 78:1248, 2010.

\bibitem{Aurell-Boffetta-Crisanti-Paladin-Vulpiani-96}
E.~Aurell, G.~Boffetta, A.~Crisanti, G.~Paladin, and A.~Vulpiani.
\newblock Growth of non-infinitesimal perturbations in turbulence.
\newblock {\em Phys. Rev. Lett.}, 77:1262--1265, 1996.

\bibitem{badiicomplexity}
R~Badii and A~Politi.
\newblock {\em Complexity hierarchical structures and scaling in physics,
  1997}.
\newblock Cambridge University Press, Cambridge, UK.

\bibitem{batterman1996chaos}
Robert~W Batterman and Homer White.
\newblock Chaos and algorithmic complexity.
\newblock {\em Foundations of Physics}, 26(3):307--336, 1996.

\bibitem{davey2010enantiomer}
S.~Davey.
\newblock Enantiomer separation: Selective soap-films.
\newblock {\em Nature Chemistry}, 2010.

\bibitem{derjaguin1989theory}
B.V. Derjaguin.
\newblock {\em {Theory of stability of colloids and thin films}}.
\newblock Consultants Bureau New York and London, 1989.

\bibitem{doering1995applied}
Charles~R Doering and John~D Gibbon.
\newblock {\em Applied analysis of the Navier-Stokes equations}, volume~12.
\newblock Cambridge University Press, 1995.

\bibitem{erneux1993nonlinear}
Thomas Erneux and Stephen~H Davis.
\newblock Nonlinear rupture of free films.
\newblock {\em Physics of Fluids A: Fluid Dynamics (1989-1993)},
  5(5):1117--1122, 1993.

\bibitem{exerowa1998foam}
D.~Exerowa and P.M. Kruglyakov.
\newblock {\em {Foam and foam films: theory, experiment, application}}.
\newblock Elsevier, New York, 1998.

\bibitem{frederickson1983liapunov}
Paul Frederickson, James~L Kaplan, Ellen~D Yorke, and James~A Yorke.
\newblock The liapunov dimension of strange attractors.
\newblock {\em Journal of Differential Equations}, 49(2):185--207, 1983.

\bibitem{israelachvili1991intermolecular}
J.N. Israelachvili.
\newblock {\em {Intermolecular and surface forces}}.
\newblock Academic press London, 1991.

\bibitem{jones1966stability}
M.N. Jones, K.J. Mysels, and P.C. Scholten.
\newblock {Stability and some properties of the second black film}.
\newblock {\em Transactions of the Faraday Society}, 62:1336--1348, 1966.

\bibitem{kantz2004nonlinear}
Holger Kantz and Thomas Schreiber.
\newblock {\em Nonlinear time series analysis}, volume~7.
\newblock Cambridge university press, 2004.

\bibitem{kellay2011turbulence}
H.~Kellay.
\newblock Turbulence: Thick puddle made thin.
\newblock {\em Nature Physics}, 7(4):279--280, 2011.

\bibitem{Kruesemann-2012}
Henning Krusemann.
\newblock {Lattice Boltzmann Simulation of 2D Thermal Convection }, 2012.

\bibitem{oron1997long}
A.~Oron, S.H. Davis, and S.G. Bankoff.
\newblock {Long-scale evolution of thin liquid films}.
\newblock {\em Reviews of Modern Physics}, 69(3):931--980, 1997.

\bibitem{ott2002chaos}
Edward Ott.
\newblock {\em Chaos in dynamical systems}.
\newblock Cambridge university press, 2002.

\bibitem{ottino1989kinematics}
JM~Ottino.
\newblock {\em The Kinematics of Mixing}.
\newblock Cambridge Univ Pr, June 1989.

\bibitem{Prudhomme-Khan-96}
Eds. Prud'homme R.~K., Khan S.~A.
\newblock {\em Foams: Theory, Measurements, and Applications}.
\newblock Dekker, N.Y., 1996.

\bibitem{reiter1998artistic}
G.~Reiter.
\newblock The artistic side of intermolecular forces.
\newblock {\em Science}, 282(5390):888, 1998.

\bibitem{ruelle1989chaotic}
David Ruelle.
\newblock {\em Chaotic evolution and strange attractors}, volume~1.
\newblock Cambridge University Press, 1989.

\bibitem{seychelles2008thermal}
F.~Seychelles, Y.~Amarouchene, M.~Bessafi, and H.~Kellay.
\newblock {Thermal Convection and Emergence of Isolated Vortices in Soap
  Bubbles}.
\newblock {\em Phys. Rev. Lett.}, 100(14):144501, 2008.

\bibitem{stoeckle2010dynamics}
S.~St\"ockle, P.~Blecua, H.~M\"ohwald, and R.~Krastev.
\newblock {Dynamics of Thinning of Foam Films Stabilized by
  n-Dodecyl-$\beta$-maltoside}.
\newblock {\em Langmuir}, 26 (7):4974--4977, 2010.

\bibitem{tel2006chaotic}
Tam{\'a}s T{\'e}l and M{\'a}rton Gruiz.
\newblock {\em Chaotic dynamics: an introduction based on classical mechanics}.
\newblock Cambridge University Press, 2006.

\bibitem{vermant2011fluid}
J.~Vermant.
\newblock Fluid mechanics: When shape matters.
\newblock {\em Nature}, 476(7360):286--287, 2011.

\bibitem{Verwey-Overbeek-48}
E.~J.~W. Verwey and J.~T.~G. Overbeek.
\newblock {\em {The Theory of the Stability of Lipophobic Colloids}}.
\newblock Elsevier, Amsterdam, 1948.

\bibitem{winkler2011droplet}
M.~Winkler and M.~Abel.
\newblock Droplet coalescence in 2d thermal convection of a thin film.
\newblock {\em Journal of Physics Conference Series}, 333(1):012018, December
  2011.

\bibitem{winkler2013mixing}
M.~Winkler and M.~Abel.
\newblock Mixing in thermal convection of very thin free-standing films.
\newblock {\em Physica Scripta Volume T}, 155(1):014020, July 2013.

\bibitem{winkler2013exponentially}
M.~{Winkler}, G.~{Kofod}, R.~{Krastev}, S.~{St{\"o}ckle}, and M.~{Abel}.
\newblock Exponentially fast thinning of nanoscale films by turbulent mixing.
\newblock {\em Physical Review Letters}, 110(9):094501, March 2013.

\bibitem{young1982dimension}
Lai-Sang Young.
\newblock Dimension, entropy and lyapunov exponents.
\newblock {\em Ergodic theory and dynamical systems}, 2(01):109--124, 1982.

\bibitem{yunker2011suppression}
P.J. Yunker, T.~Still, M.A. Lohr, and AG~Yodh.
\newblock Suppression of the coffee-ring effect by shape-dependent capillary
  interactions.
\newblock {\em Nature}, 476(7360):308--311, 2011.

\bibitem{zhang2005velocity}
J.~Zhang and XL~Wu.
\newblock {Velocity intermittency in a buoyancy subrange in a two-dimensional
  soap film convection experiment}.
\newblock {\em Physical review letters}, 94(23):234501, 2005.

\end{thebibliography}
\end{document}